\documentstyle[proc,epsf]{rspublic}

\def\thedemobiblio#1{\smallskip\par
 \list{}{\labelwidth 0pt \leftmargin 1em \itemindent -1em \itemsep 1pt}
 \small \parindent 0pt
 \parskip 1.5pt plus .1pt\relax
 \def\newblock{\hskip .11em plus .33em minus .07em}
 \sloppy\clubpenalty4000\widowpenalty4000
 \sfcode`\.=1000\relax}

\begin{document} 


\title[Geometric Quantum Mechanics]
{Geometric Quantum Mechanics} 
 
\author[D.C. Brody, L.P. Hughston]{
Dorje C. Brody$^{1}$ 
and 
Lane P. Hughston$^{2}$ 
} 
\affiliation{$^{1}$Blackett Laboratory, Imperial College, 
London SW7 2BZ, UK \\ 
and DAMTP, Silver Street, Cambridge CB3 9EW, UK \\  
$^{2}$Department of Mathematics, King's College London, \\ 
Strand, London WC2R 2LS, UK} 

\date{\today} 

\maketitle 
\input{psfig.sty}

\begin{abstract} 
The manifold of pure quantum states can be regarded as a complex 
projective space endowed with the unitary-invariant Riemannian 
geometry of Fubini and Study. According to the principles of 
geometric quantum mechanics, the detailed physical 
characteristics of a given quantum system can be represented by 
specific geometrical features that are selected and 
preferentially identified in this complex manifold. In particular, 
any specific feature of projective geometry gives rise to a 
physically realisable characteristic 
in quantum mechanics. Here we construct a number of examples of 
such geometrical features as they arise in the state spaces for 
spin-$\frac{1}{2}$, spin-1, and spin-$\frac{3}{2}$ systems, and for 
pairs of spin-$\frac{1}{2}$ systems. A study is undertaken on the 
geometry of entangled states, and a natural measure is assigned to 
the degree of entanglement of a given state for a general 
multi-particle system. The properties of this measure 
are analysed in detail for the entangled states of a pair of 
spin-$\frac{1}{2}$ particles, thus enabling us to determine the 
structure of the space of maximally entangled states. With the 
specification of a quantum Hamiltonian, the 
resulting Schr\"odinger trajectories induce an isometry of the 
Fubini-Study manifold. For a generic quantum evolution, the 
corresponding Killing trajectory is quasiergodic on a toroidal 
subspace of the energy surface. When the dynamical trajectory is 
lifted orthogonally to Hilbert space, it induces a geometric phase 
shift on the wave function. The uncertainty of an observable in a 
given state is the length of the gradient vector of the level 
surface of the expectation of the observable in that state, a 
fact that allows us to calculate higher order corrections to the 
Heisenberg relations. A general mixed state is determined by a 
probability density 
function on the state space, for which the associated first moment 
is the density matrix. The advantage of the idea of a general state 
is in its applicability in various attempts to go beyond the 
standard quantum theory, some of which admit a natural phase-space 
characterisation. 
\vskip .2cm 
\begin{center} 
{\footnotesize {\bf Keywords: Quantum phase space; projective 
geometry; \\ quantum entanglement; Kibble-Weinberg theory}} 
\end{center} 
\end{abstract} 

\vskip .5cm

\section{Introduction} 

The line of investigation which we refer to as `Geometric Quantum 
Mechanics' originated over two decades ago in the work of Kibble 
(1978, 1979), who showed how quantum theory could be formulated in 
the language of Hamiltonian phase-space dynamics. This was a 
remarkable development inasmuch as previously it was believed by 
physicists that {\it classical} mechanics has a natural Hamiltonian 
phase-space structure, to which one had to apply a suitable 
{\it quantisation procedure} to produce a very different kind of 
structure, namely, the complex Hilbert space of quantum mechanics 
together with a family of linear operators, corresponding to 
physical observables. However, with the advent of geometric quantum 
mechanics it has become difficult to sustain this point 
of view, and quantum theory has come to be recognised much more as 
a self-contained entity.   \par 

A notable attempt to codify the quantisation procedure in a 
rigourous mathematical framework was pursued in the geometric 
quantisation program (see, e.g., Woodhouse 1992). Geometric quantum 
mechanics, however, is not concerned with the 
quantisation procedure, as such, but accepts quantum 
theory as given. Indeed, from a modern perspective the 
current flows in reverse, and a major objective is to 
understand how the classical world emerges from quantum 
theory. Thus, in contrast to the aforementioned `geometric 
quantisation' program, what we really need might be more 
appropriately called a `geometric classicalisation' program. \par 

To this extent, there may even be grounds for arguing that the 
notion of quantisation is superfluous. Present thinking on these 
issues is based on a special relationship between classical and 
quantum mechanics distinct from the quantisation idea, namely, 
that quantum theory possesses an intrinsic mathematical structure 
equivalent to that of Hamiltonian phase-space dynamics, only the 
underlying phase-space is not that of classical mechanics, but 
rather the quantum mechanical state space itself, i.e., what we 
call the `space of pure states'. \par 

The approach to quantum mechanics via its natural phase-space 
geometry initiated by Kibble offers insights into many of the 
more enigmatic aspects of the theory: linear superposition of 
states, Schr\"odinger evolution, quantum entanglement, quantum 
probability laws, uncertainty relations, geometric phases, and 
the collapse of the wave function. One of the goals of this 
article is to illustrate in geometrical terms the interplay 
between these aspects of quantum theory. \par 

The plan of the paper is as follows. In \S 2-4 we introduce the 
projective geometric framework, and review the main features of 
the quantum phase space. In \S 5 the phase space of a spin-1 
system is studied, and in \S 6 we look at a spin-$\frac{3}{2}$ 
system, relating the properties of this system to the geometry 
of the twisted cubic curve in $CP^{3}$. In \S 7 we develop a 
geometric theory of entangled states and discuss the 
properties of quantum measurements made on such systems. This 
theory is extended in \S 8-10 where we introduce a new measure 
of entanglement, and explore its applications. \par 

In \S 11-14 we consider quantum dynamics from a geometric view, 
and demonstrate in particular a quasiergodic property satisfied 
by the Schr\"odinger trajectories. We show that the theory of 
geometric phase has a natural characterisation in this setting, 
thus allowing us to introduce a quantum mechanical analogue of 
the Poincare integral invariant. Then in \S 15 we examine the 
status of mixed states in the geometric 
framework of relevance to quantum statistical mechanics, and 
discuss the merits of general states characterised by density 
functions on the quantum phase space. \par 

Following the original observations of Kibble, many authors (see, 
e.g., Heslot 1985; Anandan $\&$ Aharonov 1990; Cirelli, et. al. 
1990; Gibbons 1992, 1997; Gibbons $\&$ Phole 1993; Hughston 1995, 
1996; Ashtekar $\&$ Schilling 1998; Brody $\&$ Hughston 1999a; 
Field $\&$ Hughston 1999; Adler $\&$ Horwitz 1999) have 
contributed to the further development of geometric quantum 
mechanics, and in doing so have demonstrated that this 
methodology not only provides new insights into the workings of 
the quantum world as we presently understand it, but also acts as 
a base from which extensions of standard quantum theory can be 
developed, some of which we shall touch upon briefly towards the 
end of this article, in \S 16. \par 

\section{Projective state space} 

Let us begin by reviewing briefly how quantum 
mechanics is ordinarily formulated. A physical system is 
represented by a wave function $\psi({\bf x},t)$, which for each 
time $t$ belongs to a complex Hilbert space 
${\cal H}$. We also require a set of linear operators on ${\cal H}$, 
corresponding to observables. The wave function 
characterises the `state' of the system at time $t$. 
In the case of a single particle of mass $m$ moving in Euclidean 
3-space under the influence of a potential $\phi({\bf x})$, the 
evolution of the system is given by Schr\"odinger's wave 
equation 
\[ 
{\rm i}\hbar\frac{\partial}{\partial t}\psi({\bf x},t)\ =\ 
\left( -\frac{1}{2m}\nabla^{2} + \phi({\bf x})\right) 
\psi({\bf x},t)\ . 
\] 
Given an initial condition $\psi({\bf x},0)$, the Schr\"odinger 
equation determines the development of the state, in terms of 
which we can then calculate the expectation of any observable. \par 

Physical properties of the system depend on the wave function 
only up to an overall 
complex factor. For instance, suppose we consider an observation to 
determine whether the particle lies in a region $D$ in 
${\bf R}^{3}$. We define the linear operator $\chi_{_D}$, the 
{\it characteristic function} for $D$, by the property 
$\chi_{_D}\psi({\bf x}) = \psi({\bf x})$ for 
${\bf x}\in D$ and $\chi_{_D}\psi({\bf x}) = 0$ for ${\bf x}\notin 
D$. Thus $\chi_{_D}$ `truncates' the wave function outside $D$. 
In particular, $\chi_{_D}$ has two eigenvalues, 1 and 
0, corresponding to eigenfunctions concentrated on $D$ and on 
the complement of $D$ in ${\bf R}^{3}$. 
The probability of an affirmative result for a measurement 
to determine whether the particle lies in $D$ is 
given by the expectation of the operator $\chi_{_D}$, that is, 
\[ 
{\rm E}[\chi_{_D}]\ =\ \frac{\int_{{\bf R}^{3}}{\bar\psi}({\bf x})
\chi_{_D}\psi({\bf x})d^{3}x}
{\int_{{\bf R}^{3}}{\bar\psi}({\bf x})\psi({\bf x})d^{3}x}\ . 
\] 
In this case, we note that the probability density function 
\[ 
p({\bf x})\ =\ \frac{{\bar\psi}({\bf x})\psi({\bf x})}
{\int_{{\bf R}^{3}}{\bar\psi}({\bf x})\psi({\bf x})d^{3}{\bf x}}  
\] 
on ${\bf R}^{3}$ is independent of the phase and scale of 
$\psi({\bf x})$. In other words, the state of the system is 
not given by $\psi({\bf x})$ itself, but rather by an equivalence 
class modulo transformations of the form 
\[ 
\psi({\bf x},t)\ \rightarrow\ e^{{\rm i}\lambda(t)}\psi({\bf x},t) 
\] 
for any complex time-dependent function $\lambda(t)$. For this 
reason, we say the state is given, at any time, by 
a `ray' through the origin in ${\cal H}$. The space of such rays 
is called projective Hilbert space, denoted ${\cal PH}$. All of the 
operations of quantum mechanics can be referred to ${\cal PH}$ 
directly, without consideration of ${\cal H}$ itself. For 
example, the Schr\"odinger equation is not invariant 
under a change of phase and scale for $\psi({\bf x})$, 
whereas the {\it projective} Schr\"odinger equation 
\begin{eqnarray} 
{\rm i}\hbar\left[ 
\psi({\bf y})\frac{\partial\psi({\bf x})}{\partial t} 
- \psi({\bf x})\frac{\partial\psi({\bf y})}{\partial t}\right] 
\ &=&\ -\frac{1}{2m}\left[\psi({\bf y})\nabla^{2}\psi({\bf x}) 
- \psi({\bf x})\nabla^{2}\psi({\bf y})\right] \nonumber \\ 
& &\ + \left[\phi({\bf x})-\phi({\bf y})\right]
\psi({\bf x})\psi({\bf y}) \nonumber 
\end{eqnarray} 
is, in fact, invariant under such transformations. Had 
Schr\"odinger elected to present this relation as his wave 
equation, none of the physical consequences would have 
differed. \par 

\section{Pure states} 

There is a beautiful geometry associated with the projective 
Hilbert space ${\cal PH}$ which 
is so compelling in its richness that, in our opinion, all 
physicists should become acquainted with it. The basic idea can 
be sketched as follows. For simplicity we use an index notation for 
the Hilbert space ${\cal H}$. Instead of $\psi({\bf x})$ we write 
$\psi^{\alpha}$, where the Greek index $\alpha$ labels components 
of the Hilbert-space vector with respect to a basis. This notation 
serves us equally well whether ${\cal H}$ is finite or infinite 
dimensional. The highly effective use of the index notation for 
Hilbert space was first popularised by Geroch (1970). For the 
complex conjugate of $\psi^{\alpha}$ we write ${\bar \psi}_{\alpha}$. 
The `downstairs' index reminds us that ${\bar \psi}_{\alpha}$ is 
a `bra' vector, i.e., it belongs to the dual of the vector 
space to which $\psi^{\alpha}$ belongs. \par 

\begin{figure}[b] 
\centerline{ 
\psfig{file=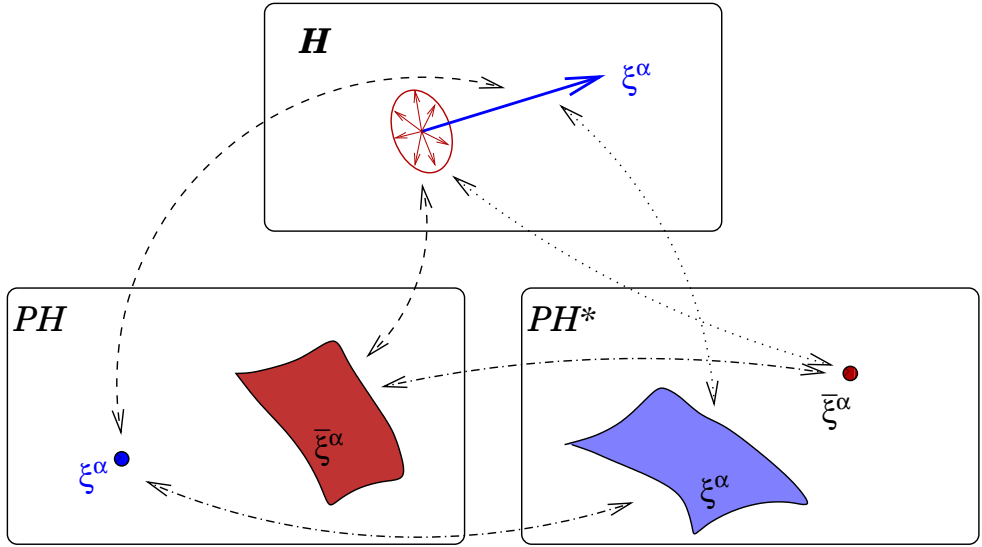,width=13cm,angle=0} 
}
 \caption{{\it Hermitian correspondence}. A pure quantum 
mechanical state corresponds to a ray through the origin $O$ 
in complex Hilbert space ${\cal H}$. Such a ray is given by a 
Hilbert space vector $\xi^{\alpha}$, specified up to 
proportionality, which can also be used as a set of `homogeneous 
coordinates' for a point in the projective Hilbert space ${\cal PH}$. 
The states $\psi^{\alpha}$ orthogonal to $\xi^{\alpha}$ constitute 
a projective hyperplane in ${\cal PH}$, with the equation 
${\bar \xi}_{\alpha}\psi^{\alpha}=0$. This hyperplane corresponds 
to a point ${\bar \xi}_{\alpha}$ in the dual projective space 
${\cal PH}^{*}$. 
} 
\end{figure} 

The usual inner product 
between $\psi^{\alpha}$ and ${\bar \psi}_{\alpha}$ can be written 
${\bar \psi}_{\alpha}\psi^{\alpha}$, with an implied summation 
over the repeated index. In the case of a wave function, this is 
equivalent to $\int_{{\bf R}^{3}}
{\bar \psi}({\bf x})\psi({\bf x})d^{3}x$, which in the Dirac 
bra-ket notation is $\langle{\bar\psi}|\psi\rangle$. 
By use of the index notation the Schr\"odinger equation can be 
represented in the compact form 
${\rm i}\hbar\partial_{t}\psi^{\alpha} = 
H^{\alpha}_{\beta}\psi^{\beta}$, where $H^{\alpha}_{\beta}$ is 
the Hamiltonian operator, $\partial_{t}=\partial/\partial t$, and 
for the projective Schr\"odinger equation we have 
\[ 
{\rm i}\hbar \psi^{[\alpha}\partial_{t}\psi^{\beta]}
\ =\ \psi^{[\alpha}H^{\beta]}_{\gamma}\psi^{\gamma} , 
\] 
where the skew brackets indicate antisymmetrisation. \par 

A Hilbert space vector $\xi^{\alpha}$ can also represent 
homogeneous coordinates for the corresponding point in the 
projective Hilbert space ${\cal PH}$. This is valid when we 
consider relations homogeneous in $\xi^{\alpha}$, for 
which the scale is irrelevant. 
For example the complex conjugate ${\bar \xi}_{\alpha}$ of a 
`point' in ${\cal PH}$ can be represented by the linear 
subspace (hyperplane) of points 
$\psi^{\alpha}$ in ${\cal PH}$ satisfying ${\bar \xi}_{\alpha}
\psi^{\alpha}=0$. The set of all such hyperplanes constitutes 
the dual space ${\cal PH}^{*}$. The points of ${\cal PH}^{*}$ 
correspond to hyperplanes in ${\cal PH}$. Conversely, the 
points of ${\cal PH}$ correspond to hyperplanes in ${\cal PH}^{*}$, 
as illustrated in Figure 1. \par 

One of the advantages of the use of 
projective geometry in the present context is that it allows us 
to represent states (points) and dual states (hyperplanes) as 
geometrical objects coexisting in the same space ${\cal PH}$. 
The complex conjugation operation, in particular, determines a 
{\it Hermitian correspondence} between points and their orthogonal 
hyperplanes. \par  

\section{Superposition of states} 

The join of two distinct points $\xi^{\alpha}$ and 
$\eta^{\alpha}$ in ${\cal PH}$ is a complex projective line, 
represented by points in ${\cal PH}$ of the form 
\[ 
\psi^{\alpha}\ =\ A\xi^{\alpha}+B\eta^{\alpha}\ , 
\] 
where $A$ and $B$ 
are complex numbers, not both zero. A neat way of characterising 
this line is the tensor $L^{\alpha\beta}=
\xi^{[\alpha}\eta^{\beta]}$. Physically, $L^{\alpha\beta}$ 
represents the system of all possible 
quantum mechanical superpositions of the states 
$\xi^{\alpha}$ and $\eta^{\alpha}$. To see that $L^{\alpha\beta}$ 
represents a line, consider a finite dimensional case where 
${\cal PH}=CP^{n}$. Then, because of the skew-symmetry of 
$L^{\alpha\beta}$ it has $\frac{1}{2}n(n+1)$ complex components, 
which can be viewed as the line coordinates of the given line. 
The fundamental property of these line coordinates is 
that their ratios are independent of the choice of the two 
points $\xi^{\alpha}$ and $\eta^{\alpha}$, in such a way that 
all points on the given line are treated on an equal footing. \par 

\begin{figure}[t] 
   \psfig{file=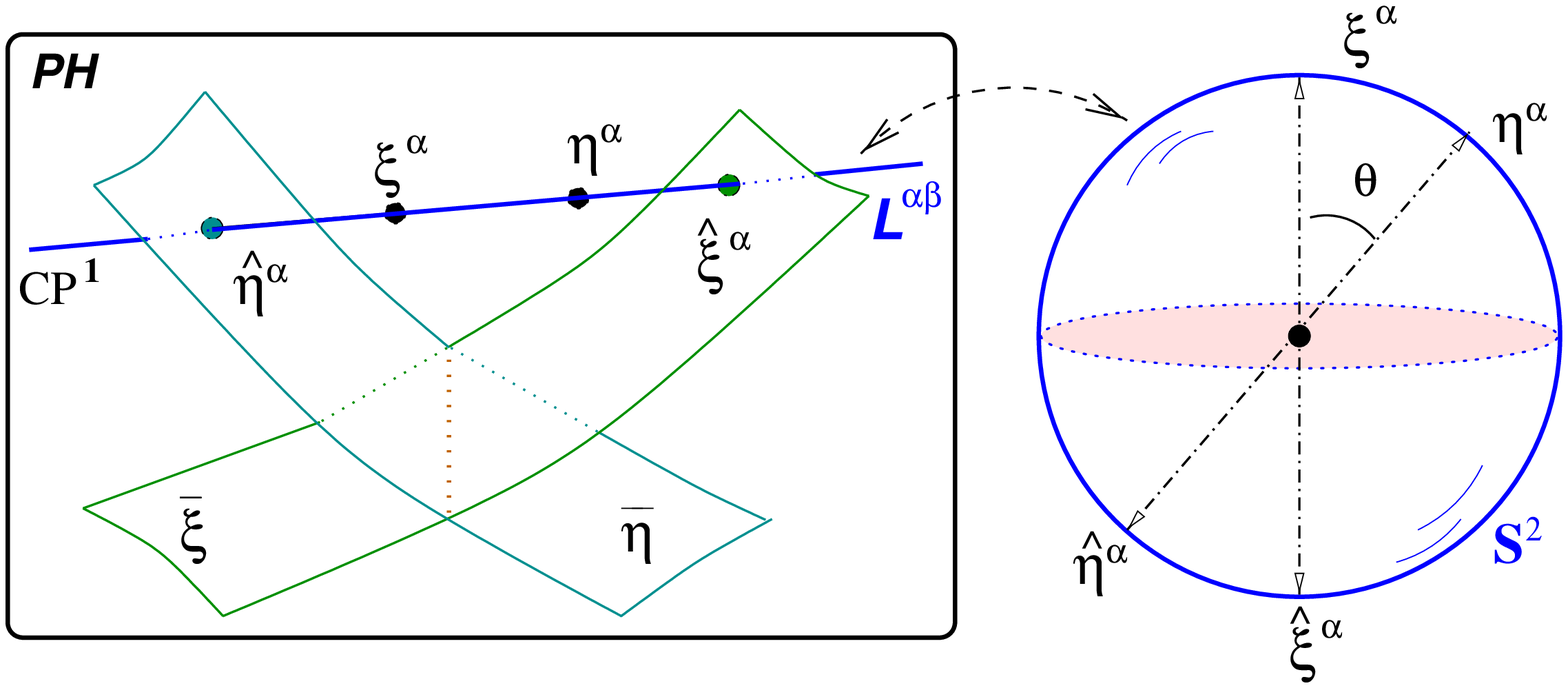,width=13cm,angle=0} 
\vskip .25cm
 \caption{{\it Transition probability}. The join of two states 
$\xi^{\alpha}$ and $\eta^{\alpha}$ 
in projective Hilbert space ${\cal PH}$ is a complex projective 
line $CP^{1}$: $L^{\alpha\beta}=\xi^{[\alpha}\eta^{\beta]}$. The 
points on $L^{\alpha\beta}$ represent superpositions 
of $\xi^{\alpha}$ and $\eta^{\alpha}$. Such a line is intrinsically 
a real 2-manifold with spherical topology. The conjugate 
hyperplanes ${\bar\xi}_{\alpha}$ and ${\bar\eta}_{\alpha}$ 
intersect $L^{\alpha\beta}$ at points ${\hat\xi}^{\alpha}$ and 
${\hat\eta}^{\alpha}$ in ${\cal PH}$. The angle $\theta$ 
determined by the cross ratio $\cos^{2}(\theta/2) = \xi^{\alpha}
{\bar\eta}_{\alpha}\eta^{\beta}{\bar\xi}_{\beta}/\xi^{\gamma}
{\bar\xi}_{\gamma}\eta^{\delta}{\bar\eta}_{\delta}$ induces a 
metrical geometry on $S^{2}$, for which $\theta$ is the usual 
angular distance, and ${\hat\xi}^{\alpha}$ is antipodal to 
$\xi^{\alpha}$. 
} 
\end{figure} 

The simplest situation in which a probabilistic idea arises in 
quantum theory is also the simplest situation in which the concept 
of the `distance' between two states arises. The transition 
probability for the states $\xi^{\alpha}$ and $\eta^{\alpha}$ 
determines an angle $\theta$ as follows: 
\[ 
\cos^{2}\frac{1}{2}\theta\ =\ \frac{
\xi^{\alpha}{\bar\eta}_{\alpha}\eta^{\beta}{\bar\xi}_{\beta}}
{\xi^{\gamma}{\bar\xi}_{\gamma}\eta^{\delta}{\bar\eta}_{\delta}}\ . 
\] 
Clearly, $\theta$ is independent of the scale and phase of 
$\xi^{\alpha}$ and $\eta^{\alpha}$. This angle defines a distance 
between the states $\xi^{\alpha}$ and $\eta^{\alpha}$ in ${\cal PH}$. 
If the states coincide, then $\theta=0$; for orthogonal states we 
have $\theta=\pi$, the maximum distance. \par 

Suppose we set $\theta=ds$ and $\xi^{\alpha}=
\psi^{\alpha}$, $\eta^{\alpha}=\psi^{\alpha}+d\psi^{\alpha}$. By 
use of the expression for the transition probability, expanded to 
second order, we find that the infinitesimal distance $ds$ 
between two neighbouring states is 
\[ 
ds^{2}\ =\ 8\frac{
\psi^{[\alpha}d\psi^{\beta]}
{\bar \psi}_{[\alpha}d{\bar \psi}_{\beta]}}
{({\bar\psi}_{\gamma}\psi^{\gamma})^{2}}\ , 
\] 
an expression known to geometers as the Fubini-Study metric 
(Kobayashi $\&$ Nomizu 1969; Arnold 1989). This expression is 
well-defined both in finite and infinite dimensions. The 
introduction of the Fubini-Study geometry illustrates how the 
notions of probability and distance become interlinked, once quantum 
theory is formulated in a geometric manner. The {\it geodesic 
distance} with respect to the Fubini-Study metric determines the 
transition probability between two states. Indeed, the nontrivial 
metrical geometry of the Fubini-Study manifold is responsible for 
the `peculiarities' of the quantum world, and in what follows 
we shall see various examples of this phenomenon. \par 

\section{Spin measurements} 

The specification of a physical system implies further geometrical 
structure on the state space. Indeed, the point of view we suggest 
is that {\it all} the relevant physical details of a quantum system 
can be represented by additional projective geometrical features. 
Here and in subsequent sections we shall illustrate this point with 
several examples. 
Let us first consider the spin degrees of freedom of a 
nonrelativistic spin-1 particle, as represented by a symmetric 
spinor $\phi^{AB}$ ($A,B=0,1$). The relevant Hilbert space 
has three dimensions, and we denote the corresponding projective 
Hilbert space $CP^{2}$. A symmetric spinor has a natural 
decomposition $\phi^{AB}=\alpha^{(A}\beta^{B)}$, where $\alpha^{A}$ 
and $\beta^{A}$ are called `principal spinors', and round 
brackets denote symmetrisation. There is a special conic 
${\cal C}$, corresponding to degenerate spinors of the form 
$\phi^{AB}=\psi^{A}\psi^{B}$ for some repeated principal spinor 
$\psi^{A}$. \par 

\begin{figure}[t] 
   \psfig{file=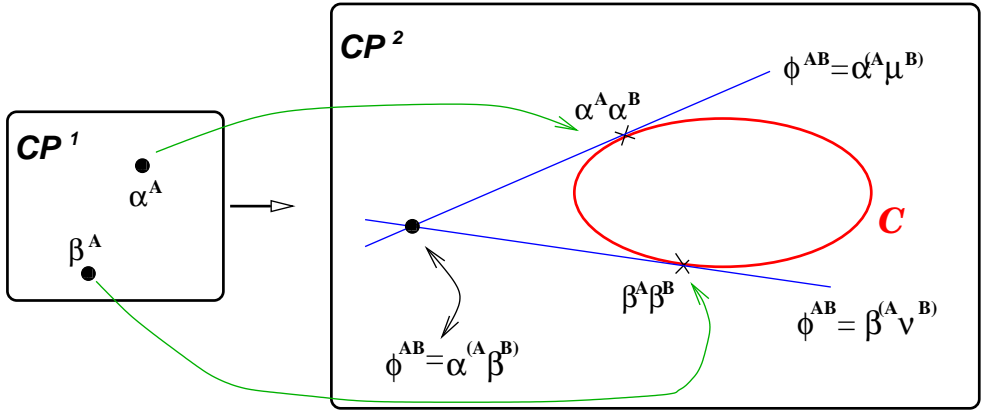,width=13cm,angle=0} 
\vskip .25cm 
 \caption{{\it Spin-{\rm 1} particle}. A symmetric spinor 
$\phi^{AB}$ has three independent components which act as 
homogeneous coordinates for $CP^{2}$. The 
image of the map ${\cal C}:\ CP^{1}\rightarrow CP^{2}$, defined by 
$\{\psi^{A}\in CP^{1}\}\rightarrow\{\psi^{A}\psi^{B}\in CP^{2}\}$ 
determines a curve ${\cal C}$ in $CP^{2}$. The tangent to 
${\cal C}$ at the point $\phi^{AB}=\alpha^{A}\alpha^{B}$ in 
$CP^{2}$ consists of spinors of the form $\phi^{AB}=\alpha^{(A}
\mu^{B)}$ for some $\mu^{A}$. The intersection of the lines tangent 
to the points $\alpha^{A}\alpha^{B}$ and $\beta^{A}\beta^{B}$ is the 
point $\alpha^{(A}\beta^{B)}$. Conversely, once a conic ${\cal C}$ 
is specified, a map ${\cal C}^{-1}$ is established from $CP^{2}$ 
to point-pairs in $CP^{1}$, called principal spinors. The points 
on ${\cal C}$ map to degenerate point-pairs. An analogous result 
holds for spin 2, which has striking applications in gravitational 
theory, first systematically explored by Penrose (1960). 
} 
\end{figure} 

The identification of ${\cal C}$ is sufficient to 
induce the structure of a spin-1 system on the state space, since 
through any generic point in $CP^{2}$ there are two lines tangent 
to ${\cal C}$, and the corresponding tangent points determine the 
principal spinors, up to scale, as shown in Figure 3. Alternatively, 
we can think of a conic ${\cal C}$ in $CP^{2}$ being represented 
by a map (see, e.g., Semple $\&$ Kneebone 1952) from $CP^{1}$ to $
CP^{2}$ such that if $(t,u)$ are homogeneous coordinates on $CP^{1}$, 
we have 
\[ 
{\cal C}:\  (t,u) \rightarrow (t^{2}, tu, u^{2})\ , 
\] 
where $(t^{2}, tu, u^{2})$ now represents homogeneous coordinates 
on $CP^{2}$. Because a complex projective line, in real dimensions, 
represents a sphere $S^{2}$ (cf. Figure 2), the specification of 
the spin direction determines a point on $S^{2}$, and hence on 
${\cal C}$. \par 
 
The conic is required to be compatible with 
the complex conjugation operation on the state space in the sense 
that if we conjugate a point of ${\cal C}$, then the resulting line 
is tangent to ${\cal C}$. The complex conjugate 
${\bar\phi}_{AB}={\bar\alpha}_{(A}
{\bar\beta}_{B)}$ of a general state corresponds to a complex 
projective line consisting of states of the form 
$P{\bar\alpha}^{A}{\bar\alpha}^{B} + 
Q{\bar\beta}^{A}{\bar\beta}^{B}$, where we define 
${\bar\alpha}^{A}=\epsilon^{AB}{\bar\alpha}_{B}$ and 
${\bar\beta}^{A}=\epsilon^{AB}{\bar\beta}_{B}$, with $\epsilon^{AB}$ 
the natural symplectic structure. The rules for the complex 
conjugation map {\bf c} on spinors are as follows: 
\[ 
\left\{ \begin{array}{l} 
{\bf c}(\alpha^{A})\ =\ {\bar\alpha}_{A} \\ 
{\bf c}({\bar\alpha}^{A})\ =\ -\alpha_{A} . 
\end{array} \right. 
\] 
The latter identity arises since ${\bf c}({\bar\alpha}^{A}) = 
{\bf c}(\epsilon^{AB}{\bar\alpha}_{B}) = 
\epsilon_{AB}\alpha^{B}=-\alpha_{A}$. Recall that for any spinor 
$\phi^{A}$ we have the relation $\phi^{A}=\epsilon^{AB}\phi_{B}$ 
and $\phi^{A}\epsilon_{AB}=\phi_{B}$, and that $\epsilon_{AB}$ 
satisfies $\epsilon_{AB}=-\epsilon_{BA}$ and 
$\epsilon_{AB}={\bar\epsilon}_{BA}$. If we take the complex 
conjugate of a state on ${\cal C}$, the resulting line is tangent 
to the conic at a point, which we call the conjugate of the 
original point on ${\cal C}$. This establishes a Hermitian 
correspondence between pairs of points on ${\cal C}$. 
For a state $\phi^{AB}=\psi^{A}\psi^{B}$ the conjugate line 
consists of states of the form 
$\lambda^{(A}{\bar\psi}^{B)}$ for arbitrary 
$\lambda^{A}$. This line touches the conic ${\cal C}$ at the 
point ${\bar\psi}^{A}{\bar\psi}^{B}$. \par
   
\begin{figure}[t] 
   \psfig{file=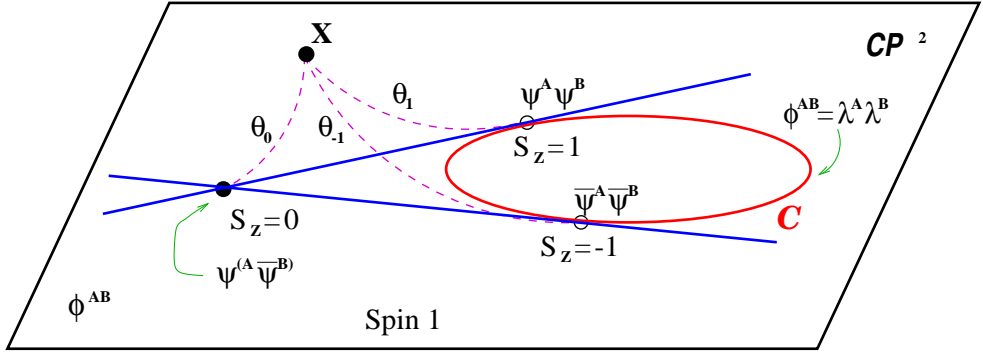,width=13cm,angle=0} 
\vskip .25cm 
 \caption{{\it Spin measurement}. The state space of a spin 1 
system has a conjugation relation that associates to each 
point $\psi^{A}\psi^{B}$ on the special conic a conjugate point 
${\bar\psi}^{A}{\bar\psi}^{B}$. The antipodal points $\psi^{A}$ 
and ${\bar\psi}^{A}$ on the corresponding 2-sphere select a 
direction in Euclidean 3-space. The three points $\psi^{A}\psi^{B}$, 
${\bar\psi}^{A}{\bar\psi}^{B}$, and $\psi^{(A}{\bar \psi}^{B)}$ are 
eigenstates of the spin operator $S_{z}$ associated with this axis. 
The corresponding geodesic distances $\theta_{1}$, $\theta_{-1}$, 
$\theta_{0}$ to a generic state ${\bf X}\in CP^{2}$ determine the 
probabilities of the measurement outcomes for $S_{z}$ for a 
particle in the state ${\bf X}$.  
} 
\end{figure} 

Each choice of a point on ${\cal C}$, as noted above, 
determines a spin axis. For any spin axis there are three 
possible spin states, with eigenvalues $1,-1$ and $0$. The 
spin eigenstates are the points 
$\psi^{A}\psi^{B}$ and ${\bar\psi}^{A}{\bar\psi}^{B}$ on 
${\cal C}$, having the eigenvalues 1 and $-1$, together with a third 
point $\psi^{(A}{\bar\psi}^{B)}$ obtained by intersecting the 
lines tangent to the conic ${\cal C}$ at the other two points, 
corresponding to eigenvalue 0, as indicated in Figure 4. \par 

When a spin measurement is made, the initial state corresponds to a 
generic point ${\bf X}$ in $CP^{2}$, and the measurement is defined 
by a spin axis. The state then `jumps' from its 
initial point to one of the three spin eigenstates associated with 
the choice of axis. Quantum theory, as such, states nothing about 
the ``mechanism'' whereby this jump is achieved. 
We can, however, compute the probabilities, and describe the result 
in geometrical terms. First we calculate the distance from ${\bf X}$ 
to each of the three spin eigenstates, by use of the Fubini-Study 
metric. This gives us three angles $\theta_{1}$, $\theta_{-1}$, and 
$\theta_{0}$. For each angle we compute $P(\theta) = \frac{1}{2}
(1+\cos\theta)$, which gives us the probability of transition to that 
particular state. It is not obvious that the three probabilities 
computed in this way sum up to one, given any initial state in 
which the measurement is performed, but they do: this is 
a `miracle' of the Fubini-Study geometry. \par 

\section{Spin-$\frac{3}{2}$ and the twisted cubic} 

We have seen that in the case of a projective plane, there is a 
conic ${\cal C}$, corresponding to degenerate spinors obtained by 
a special map from a projective line to a plane. On the other hand, 
in three-dimensional projective space $CP^{3}$ there are two 
different kinds of locus to be considered, each of which is in 
some respects a proper analogue of the conic, namely, the 
quadric surface $Q$ and the twisted cubic curve ${\cal T}$. While 
a surface is the locus of a variable point of space which has two 
complex degree of freedom, a curve is the locus of a variable point 
of space of one complex degree of freedom. When viewed as the 
state space of a quantum mechanical system, the quadric surface 
in $CP^{3}$ characterises the disentangled states of a pair of 
spin-$\frac{1}{2}$ particles, the geometry of which we shall 
study in some detail in subsequent sections. \par 

The twisted cubic, the simplest nonplanar curve in projective 
geometry, on the other hand, plays an essential role in the 
geometry of the state space of a spin-$\frac{3}{2}$ particle. 
Analogous to the conic curve, the twisted cubic can be represented 
by a map from $CP^{1}$ to $CP^{3}$ of the form 
\[ 
{\cal T}:\  (t,u) \rightarrow (t^{3},t^{2}u,tu^{2},u^{3}) \ , 
\] 
where $(t^{3},t^{2}u,tu^{2},u^{3})$ represents homogeneous 
coordinates of points on ${\cal T}$ in $CP^{3}$. If follows that 
${\cal T}$ is an algebraic space of the third order, which meets 
a generic plane of $CP^{3}$ in three points. \par 

\begin{figure}[b] 
   \psfig{file=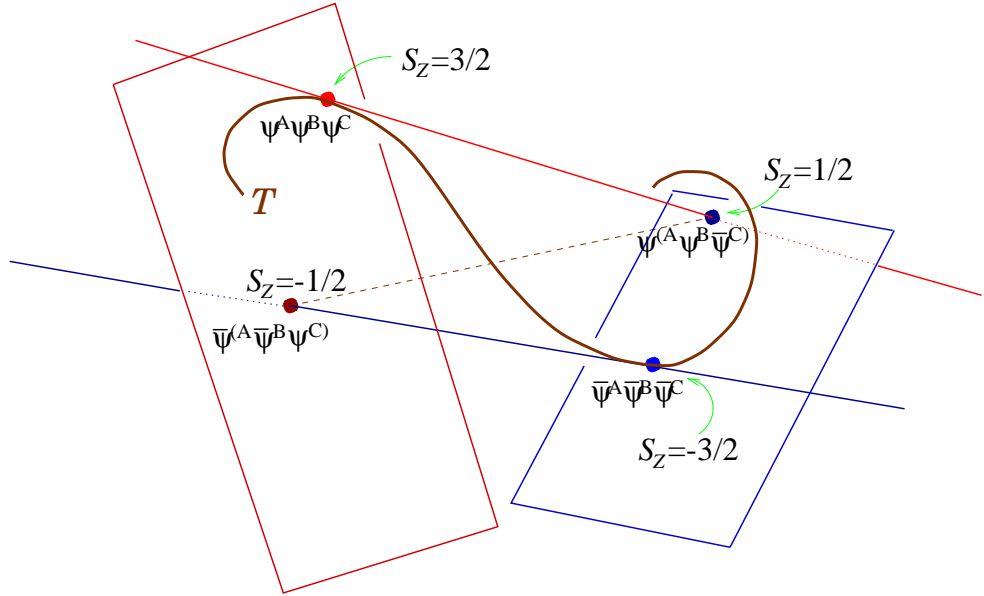,width=13cm,angle=0} 
\vskip .25cm 
 \caption{{\it The twisted cubic curve as a system of spin states}. 
The quantum phase space for a spin 3/2 particle contains a 
preferred twisted cubic ${\cal T}$ 
which is self-conjugate in the sense that the complex conjugate 
plane corresponding to any point on ${\cal T}$ necessarily 
osculates the curve at some other point on ${\cal T}$. The points 
of ${\cal T}$ are those states which have an eigenstate of 
spin 3/2 relative to some choice of spin axis. Each point of 
${\cal T}$ corresponds to a choice of spin axis and direction. 
} 
\end{figure} 

In order to proceed further, we introduce a spinorial notation 
and let the symmetric spinor $\psi^{ABC}=\psi^{(ABC)}$ denote 
homogeneous coordinates on $CP^{3}$. Then, the twisted cubic 
arising naturally here is determined by the relation 
\[ 
\tau_{AB}\ :=\ \psi_{CD(A}\psi^{CD}_{B)}\ =\ 0\ , 
\] 
where the indices on $\psi^{ABC}$ are raised and lowered according 
to the standard conventions, so for example, $\psi_{B}^{CD} = 
\epsilon_{AB}\psi^{ACD}$, and so on. The general solution to the 
algebraic relations given by $\tau_{AB}=0$ takes the form 
$\psi^{ABC}=\xi^{A}\xi^{B}\xi^{C}$ for arbitrary $\xi^{A}$ 
(Hughston, Hurd $\&$ Eastwood 1979). \par 

The specification of a twisted cubic ${\cal T}$ in $CP^{3}$ induces 
a {\it null polarity} on the state space, i.e., a natural 
correspondence between points and planes such that the polar plane 
of a given point actually includes that point. The null polarity is 
given by the map 
\[ 
\psi^{ABC}\ \rightarrow\ \psi_{ABC} = \epsilon_{AP}
\epsilon_{BQ}\epsilon_{CR}\psi^{PQR}\ , 
\] 
and it follows as an elementary spinor identity that $\psi^{ABC}
\psi_{ABC}=0$ for any choice of $\psi^{ABC}$. In the case of a 
point $\psi^{ABC}=\xi^{A}\xi^{B}\xi^{C}$ on ${\cal T}$, the 
corresponding polar plane intersects ${\cal T}$ solely at that 
point, with a three-fold degeneracy, and is called the 
{\it osculating plane} at that point. \par 

Each choice of $\xi^{A}$, 
i.e., a point on ${\cal T}$, determines a spin axis. For each spin 
axis, there are four possible spin eigenstates with eigenvalues 
$\frac{3}{2}$, $\frac{1}{2}$, $-\frac{1}{2}$, and $-\frac{3}{2}$. 
Two of the spin states, corresponding to eigenvalues 
$\pm\frac{3}{2}$, lie on ${\cal T}$ itself. These two states can 
be written $\psi^{A}\psi^{B}\psi^{C}$ and 
${\bar\psi}^{A}{\bar\psi}^{B}{\bar\psi}^{C}$, where 
${\bar\psi}^{A}=\epsilon^{AB}{\bar\psi}_{B}$ and 
${\bar\psi}_{B}={\bf c}(\psi^{B})$. \par 

The choice of $\psi^{A}$ determines the spin axis. The complex 
conjugate of the state 
$\psi^{ABC}=\psi^{A}\psi^{B}\psi^{C}$ on the twisted cubic 
${\cal T}$ is the plane 
${\bar\psi}_{ABC}={\bar\psi}_{A}{\bar\psi}_{B}{\bar\psi}_{C}$ 
in $CP^{3}$, and this plane is tangent to ${\cal T}$ at the point 
${\bar\psi}^{A}{\bar\psi}^{B}{\bar\psi}^{C}$. On the other hand, 
through the point $\psi^{A}\psi^{B}\psi^{C}$ there is a unique 
line tangent to ${\cal T}$, and this line intersects the plane 
${\bar\psi}_{A}{\bar\psi}_{B}{\bar\psi}_{C}$ at a point, given 
by $\psi^{(A}\psi^{B}{\bar\psi}^{C)}$. This point is the spin 
$\frac{1}{2}$ eigenstate with respect to that choice of axis. 
Conversely, the tangent line to ${\cal T}$ at the spin 
$-\frac{3}{2}$ state ${\bar\psi}^{A}{\bar\psi}^{B}{\bar\psi}^{C}$ 
intersects the tangent plane of ${\cal T}$ at 
$\psi^{A}\psi^{B}\psi^{C}$ at the point 
${\bar\psi}^{(A}{\bar\psi}^{B}\psi^{C)}$, which is the spin 
$-\frac{1}{2}$ state. \par 

An interesting feature of the twisted cubic geometry arises from 
the fact that for any symmetric spinor $\psi^{ABC}$ we have the 
relation 
\[ 
\tau_{AB}\psi^{ABC}\ =\ 0\ , 
\] 
which follows from the spinor identity 
$\epsilon_{[AB}\epsilon_{C]D}=0$. This relation implies that 
through any point $\psi^{ABC}$ in $CP^{3}-{\cal T}$, i.e., a point 
off the curve, there exists a unique chord of ${\cal T}$. This 
follows from the fact that, providing $\tau_{AB}$ is nondegenerate, 
the condition  $\tau_{AB}\psi^{ABC}=0$ implies a relation of the form 
\[ 
\psi^{ABC}\ =\ u\xi^{A}\xi^{B}\xi^{C} + v\eta^{A}\eta^{B}\eta^{C} 
\] 
for some $\xi^{A}$ and $\eta^{A}$ corresponding to a pair of 
spin axes such that $\xi_{A}\eta^{A}\neq0$, where $(u,v)$ 
are homogeneous coordinates on $CP^{1}$. It follows that an 
arbitrary quantum state $\psi^{ABC}$ in $CP^{3}-{\cal T}$ admits 
a unique characterisation in terms of a superposition of a pair 
of spin-$\frac{3}{2}$ eigenstates corresponding to distinct spin 
axes. If $\tau_{AB}$ is degenerate, then the chord reduces to a 
{\it tangent line} to ${\cal T}$ with a double point at the 
intersection, and $\psi^{ABC}$ has a unique representation of the 
form $\psi^{ABC}=\xi^{(A}\xi^{B}\eta^{C)}$. \par 

A similar analysis can be pursued in connection with the geometry 
of a spin-2 system, for which the state space is $CP^{4}$, endowed 
with a self-conjugate rational quartic curve. The geometry of this 
curve is closely related to the Petrov-Pirani classification of 
gravitational fields. \par 

\section{Geometry of entanglement} 

Now consider a more elaborate set-up: the spin degrees of freedom 
of an entangled pair of spin-$\frac{1}{2}$ particles. The generic 
two-particle 
state $\psi^{AB}$ for a pair of such particles (e.g., an electron 
and a positron) has a 4-dimensional Hilbert space, and the state 
space is $CP^{3}$. There is a preferred point $Z$ in $CP^{3}$, 
corresponding to the singlet state of total spin 0, for which 
$\psi^{AB} = \psi^{[AB]}$. The conjugate plane ${\bar Z}$ contains 
the triplet states of total spin 1, for which $\psi^{AB} = 
\psi^{(AB)}$. We note that ${\bar Z}$ is endowed with a conic 
${\cal C}$, each point of which defines a spin axis. There 
is also a special surface $Q\in CP^{3}$, given by the quadratic 
equation 
\[ 
\epsilon_{AC}\epsilon_{BD}\psi^{AB}\psi^{CD}\ =\ 0\ , 
\] 
consisting of states of the $disentangled$ form $\psi^{AB} = 
\xi^{A}\eta^{B}$, representing an embedding of the product of the 
state spaces of the individual spin-$\frac{1}{2}$ particles. The 
states off the quadric are the $entangled$ states. \par 

Suppose we start with a combined state of total spin 0 for the two 
particles, and we measure the spin of the first particle (say, the 
electron) relative to a given choice of axis. This  will 
disentangle the state, so the result lies on $Q$. The 
choice of axis and orientation determines a point and its 
conjugate on the conic ${\cal C}$. The tangents to the conic at 
these points intersect to form a third point off the quadric but 
in the plane of total spin 1, corresponding to a triplet state of 
eigenvalue 0 relative to the axis. We join that state to the 
starting state $Z$, and the resulting line intersects $Q$ at a 
pair of points, as shown in Figure 6. \par 

\begin{figure}[t] 
   \psfig{file=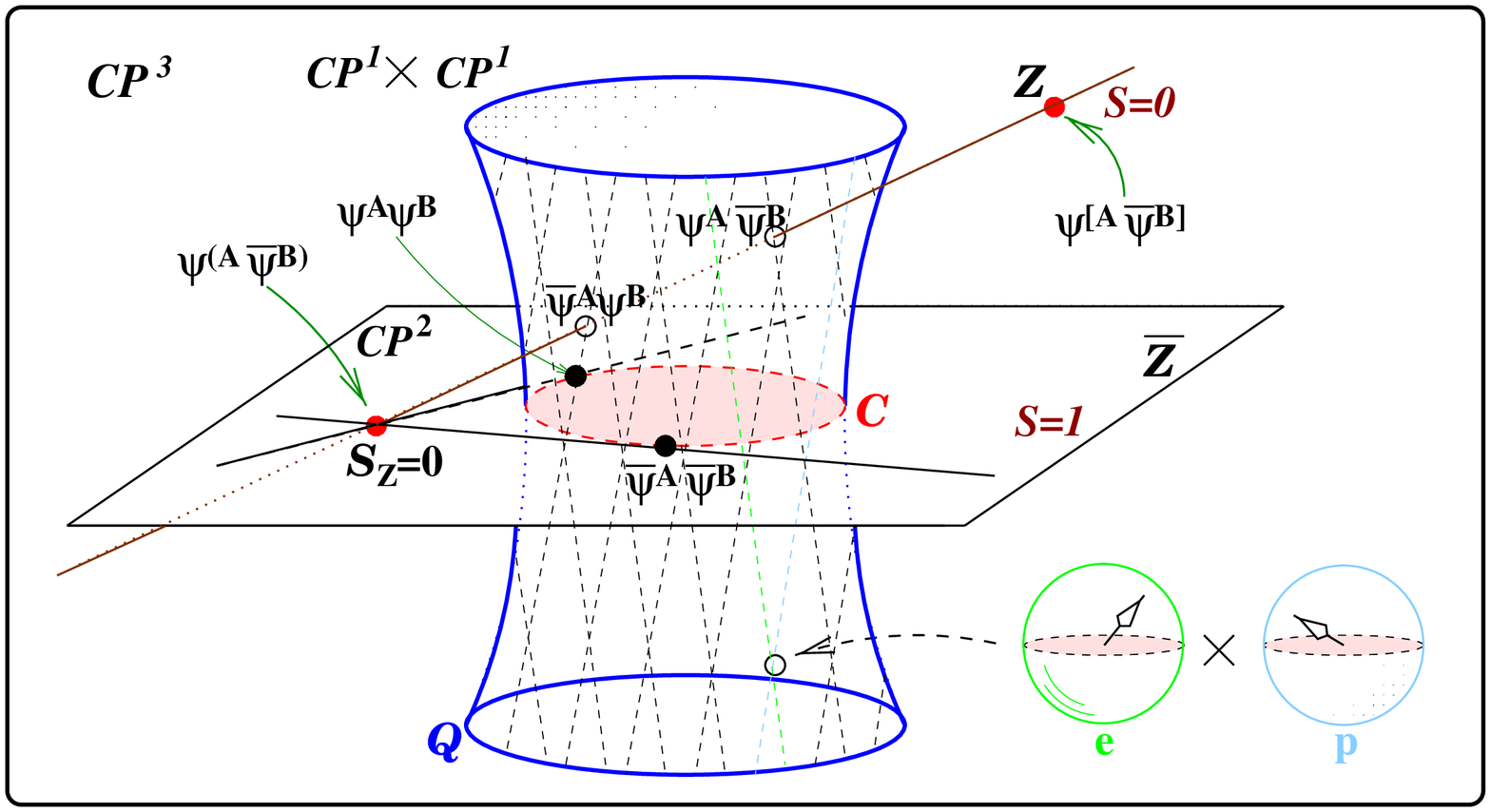,width=13cm,angle=0} 
\vskip .25cm 
 \caption{{\it Quantum entanglement}. The quantum phase space of 
an electron-positron system contains a point $Z$ for total spin 
0, and a projective hyperplane ${\bar Z}$ for total spin 1. The 
disentangled states have indefinite total spin, and comprise a 
quadric $Q$ ruled by two systems (electron and positron) of 
linear generators. 
Once a spin axis is chosen, the join of $Z$ with the state of 
total spin 1 and $S_{z}=0$ intersects $Q$ in a pair of points, 
corresponding to the possible measurement outcomes of the spin of 
the electron relative to the axis. 
} 
\end{figure} 

The two disentangled states thus obtained 
represent the possible measurement outcomes. The quadric $Q$ 
has two systems of generators, corresponding to 
the electron and positron state spaces. Through each point of 
$Q$ there is a unique `electron generator' and a unique `positron 
generator'. An electron generator represents a fixed electron state, 
each point on it corresponding to a possible positron state. The 
two points constituting the possible outcomes of the spin measurement 
of the electron have the property that their electron generators 
hit respectively the two chosen points on the conic that define the 
spin axis. The measurement result for which the electron generator 
hits the spin up state on the conic is the `electron spin up and 
positron spin down' outcome, whereas the other one is the `electron 
spin down and positron spin up' outcome. \par 

In a more general situation, the idea of quantum entanglement is 
characterised geometrically by the fact that complex projective 
space admits a Segre embedding of the form 
\[ 
CP^{m}\times CP^{n}\ \hookrightarrow\ CP^{(m+1)(n+1)-1}\ . 
\] 
Here we regard both $CP^{m}$ and $CP^{n}$ as representing the 
state space of two subsystems, respectively, while 
$CP^{(m+1)(n+1)-1}$ represents the state space of the combined 
system. One can argue that this is the main feature of 
quantum mechanics that has no analogue is classical physics. 
That is to say, classically, the state space of a combined 
system is given by the product of the state spaces of the 
subsystems, which has a moderate dimensionality when compared 
with the situation of the quantum state space for 
a combined system. \par 

\section{Measure of entanglement} 

The set up indicated above suggests a methodology according to 
which a measure $\delta(\psi)$ can be assigned to the {\it 
degree of entanglement} exhibited by a given pure state $\psi$. 
This is a topic currently of great interest in quantum physics 
(cf. Linden, et. al. 1999). Let us consider the general case of 
a finite dimensional two-particle state space $CP^{n}$ containing 
a subvariety $V^{m}\subset CP^{n}$, where 
$V^{m}=CP^{j}\times CP^{k}$ and $n=(j+1)(k+1)-1$. The variety 
$V^{m}$ represents the disentangled states of the two particles, 
and is given by the product of the respective single particle 
state space $CP^{j}$ and $CP^{k}$. \par 

We propose, as a measure of entanglement for a generic pure state 
$\psi\in CP^{n}$, the use of the {\it geodesic distance from the 
given state $\psi$ to the nearest disentangled state}. The 
distance $\delta$ is measured with respect to the 
Fubini-Study metric. \par 

Clearly, $\delta$ is a {\it natural} measure 
in the sense that it depends only on the 
Segre embedding of the variety $V^{m}$ and no additional structure 
apart from the given metrical geometry of $CP^{n}$. Furthermore, 
we can demonstrate that $\delta$ is invariant under any unitary 
transformation of $CP^{n}$ that is also an automorphism of $V^{m}$, 
i.e., transformations that preserve the disentangled state space. 
It should be evident that essentially the same construction 
applies to the case of entangled states of any number of 
particles. We do not require that the particles are necessarily 
of the same type. \par 

As a specific illustration, we consider the system described 
in \S 7 consisting of two spin-$\frac{1}{2}$ particles, where the 
state space is $CP^{3}$ and the 
space $V^{2}$ of disentangled states is a quadric 
$Q\subset CP^{3}$. Suppose we write $\psi^{AB}$ for a generic 
state, and $\bar{\psi}_{AB}$ for the corresponding complex 
conjugate hyperplane. Then the distance $\delta$ from $\psi$ to $Q$ 
is determined by the relation $\kappa=\frac{1}{2}(1+\cos\delta)$, 
where $\kappa$ is the cross ratio 
\[ 
\kappa\ =\ \frac{(\psi^{AB}{\bar X}_{AB})(X^{CD}{\bar \psi}_{CD})}
{(\psi^{AB}{\bar \psi}_{AB})(X^{CD}{\bar X}_{CD})}, 
\] 
and $X^{AB}\in Q$ maximises $\kappa$ for the given state 
$\psi^{AB}$. The cross ratio $\kappa$ is the Dirac transition 
probability from the state $\psi^{AB}$ to the state $X^{AB}$, 
and our goal is to find the states on $Q$ for which 
the transition probability from $\psi^{AB}$ is maximal. \par 

We shall turn to the details of the maximisation problem in a 
moment, in \S 10, since these are of interest, but here first we 
present the solution. Let us write 
$\psi_{CD}:= \epsilon_{AC}\epsilon_{BD}\psi^{AB}$ and 
${\bar\psi}^{AB}:= \epsilon^{AC}\epsilon^{BD}{\bar\psi}_{AB}$, 
where the antisymmetric spinor $\epsilon_{AB}$ satisfies the 
usual relation 
$\epsilon_{AB}\epsilon^{AC}=\delta^{C}_{B}$. Then the solution 
for $\kappa$ is given by $\kappa=\frac{1}{2}(1+\gamma)$, with 
\[ 
\gamma\ =\ \sqrt{1-\frac{(\psi^{AB}\psi_{AB})({\bar\psi}^{CD}
{\bar\psi}_{CD})}{(\psi^{AB}{\bar\psi}_{AB})^{2}}} . 
\] 
We note that $\gamma$ is independent of the scale of $\epsilon_{AB}$ 
and lies in the range $0\leq\gamma\leq1$. The inequality satisfied 
by $\gamma$ follows from a general result that for any element 
${\bf z}$ in a complex vector space with a Hermitian inner product 
we have the Hermitian inequality $({\bf z}\cdot{\bar{\bf z}})^{2} 
\geq ({\bf z}\cdot{\bf z})({\bar{\bf z}}\cdot{\bar{\bf z}})$. This 
can be seen by writing 
${\bf z}={\bf a}+{\rm i}{\bf b}$, where ${\bf a}$ and ${\bf b}$ 
are real, and then checking that the purported relation reduces to 
the Schwartz inequality $({\bf a}\cdot{\bf b})^{2}\leq({\bf a}\cdot
{\bf a})({\bf b}\cdot{\bf b})$. \par 

If the point $\psi^{AB}$ lies on the quadric $Q$, we have 
$\psi_{AB}\psi^{AB}=0$, and hence $\gamma=1$, which implies 
$\kappa=1$, from which it follows that the distance to the quadric 
is $\delta=0$. On the other hand, for a maximally entangled state 
the inequality is saturated at $\gamma=0$, and thus gives 
$\kappa=1/2$, which implies $\delta=\pi/2$. \par 

The interpretation of 
this result is as follows. We recall that for orthogonal states 
the Fubini-Study distance is $\pi$, the greatest distance possible 
for two states. On the other hand, the maximum distance an entangled 
state can have from the closest disentangled state, in the case of 
two spin-$\frac{1}{2}$ particles, is $\pi/2$. For example, 
with respect to a given choice of spin axis, the spin 0 singlet 
state $\epsilon^{AB}$ can be expressed as an 
antisymmetric superposition of two disentangled states, i.e., 
an up-down state and a down-up state. The two disentangled states 
are mutually orthogonal, and the singlet state lies `half way' 
between them. \par 

\section{Hermitian polar conjugation} 

Now, let us consider the geometry of this situation in more 
detail. There is a well-known construction in algebraic geometry 
according to which a proper quadric locus in $CP^{3}$ induces a 
{\it polarity} on this space --- a one-to-one correspondence 
between points and planes. Reverting briefly to the notation of 
\S 3, let us write $\psi^{\alpha}$ for the homogeneous coordinates 
of a point in $CP^{3}$, and $Q_{\alpha\beta}\psi^{\alpha}
\psi^{\beta}=0$ for the quadric. We assume that the quadric is 
nondegenerate, i.e., proper, in the sense that 
$\det(Q_{\alpha\beta})\neq0$. Then for any state $\xi^{\alpha}\in 
CP^{3}$ it follows that ${\tilde\xi}_{\alpha}:=Q_{\alpha\beta}
\xi^{\beta}$ is nonvanishing. The locus 
${\tilde\xi}_{\alpha}\psi^{\alpha}=0$ defines the {\it polar 
plane} of the point $\xi^{\alpha}$ with respect to the quadric 
$Q_{\alpha\beta}$. Since $Q_{\alpha\beta}$ is nondegenerate, 
there is a unique inverse $Q^{\alpha\beta}$ satisfying 
$Q_{\alpha\gamma}Q^{\gamma\beta}=\delta_{\alpha}^{\beta}$, and 
thus for any plane $\eta_{\alpha}$ in $CP^{3}$ we can define a 
polar point ${\tilde\eta}^{\alpha}:=Q^{\alpha\beta}\eta_{\beta}$. 
The operation is involutory in the sense that the polar 
point of the polar plane of a given point is that point. \par 

\begin{figure}[t] 
   \psfig{file=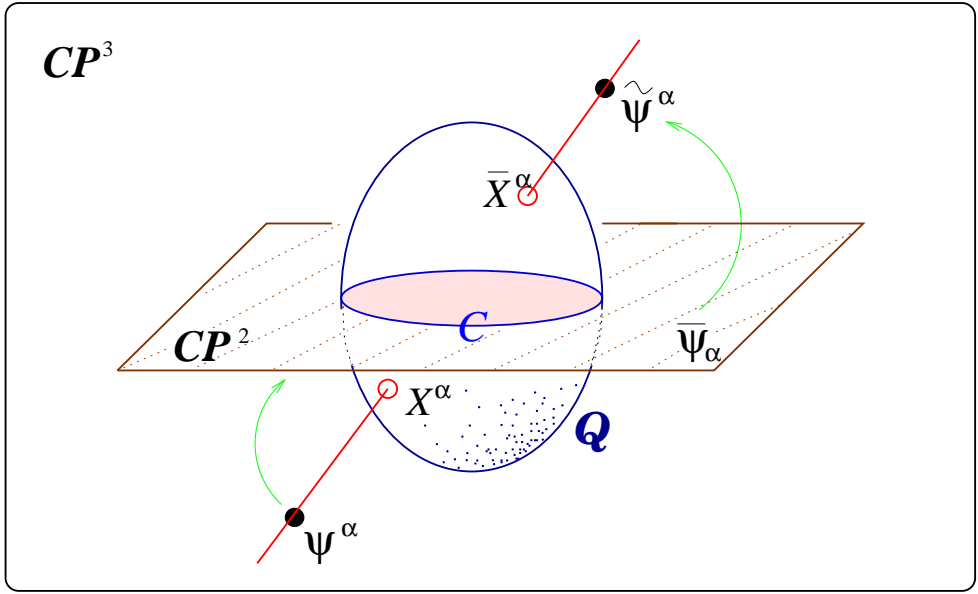,width=13cm,angle=0} 
\vskip .25cm 
 \caption{{\it Hermitian polar conjugation and the construction 
of extremal disentangled states}. Given any entangled state 
$\psi^{\alpha}$ we can form another state ${\tilde\psi}^{\alpha}$ 
given by the harmonic conjugate of the 
complex conjugate plane of $\psi^{\alpha}$ with respect to the 
quadric $Q$. The points on $Q$ nearest to and furthest from 
$\psi^{\alpha}$ are given by the intersection points $X^{\alpha}$ 
and ${\bar X}^{\alpha}$ of $Q$ with the line joining 
$\psi^{\alpha}$ and ${\tilde\psi}^{\alpha}$. 
} 
\end{figure} 

One way of constructing the polar plane of a point $\xi$ is as 
follows. Let $L$ be an arbitrary line $L^{\alpha\beta} = 
\xi^{[\alpha}\zeta^{\beta]}$ through $\xi$. Then $L$ intersects 
$Q$ twice, say, at points $A$ and $B$. Now suppose we consider 
the harmonic conjugate $\xi^{*}$ of $\xi$, on the line $L$, 
with respect to the points $A$ and $B$. This is the unique 
point $\xi^{*}$ on $L$ such that we have 
\[ 
\{\xi,\xi^{*};A,B\}\ =\ -1 
\] 
for the cross ratio. Then, as we vary $L$, the locus of 
$\xi^{*}$ sweeps out a plane, and this is the 
polar plane ${\tilde\xi}$. The polar plane ${\tilde\xi}$ 
intersects $Q$ in a conic ${\cal C}$ with the property that any 
line drawn from $\xi$ to ${\cal C}$ touches $Q$ tangentially. \par 

Conversely, if we consider all the lines through $\xi$ that 
touch $Q$ tangentially, then the union of the intersection 
points is the conic ${\cal C}$, which lies in a unique plane, the 
polar plane ${\tilde\xi}$. A point lies on its polar plane iff 
the point itself lies on the quadric, in which case the polar 
plane of the point is the tangent plane at that point. In that 
case, the conic ${\cal C}$ degenerates into a pair of lines, given 
by the two generators of the quadric through the given point. \par 

In the quantum mechanical situation we require further that the 
quadric $Q_{\alpha\beta}$ be Hermitian in the sense that 
$Q_{\alpha\beta}={\bar Q}_{\alpha\beta}$ and 
$Q^{\alpha\beta}={\bar Q}^{\alpha\beta}$. This ensures that the 
complex conjugate ket-vector $|{\bar{\tilde x}}\rangle$ of the 
polar bra-vector $\langle{\tilde x}|$ of a given ket-vector 
$|x\rangle$ agrees with the polar ket-vector 
$|{\tilde{\bar x}}\rangle$ of the complex conjugate bra-vector 
$\langle{\bar x}|$ of the given ket-vector $|x\rangle$. It follows 
that complex conjugate ket-vector of the polar bra-vector of a 
disentangled state is also disentangled, and that the polar 
ket-vector of the complex conjugate bra-vector of a disentangled 
state is disentangled. \par 

The situation described so far applies to the consideration of 
any pair of two-state systems, whether or not these systems are 
of the same type. For example, we might consider a simple toy 
model in which a lepton is regarded as a composite consisting of 
a neutral spin-$\frac{1}{2}$ particle and a spin-0 flavour doublet 
that determines whether the lepton is an electron or a muon. Then 
one might explore the properties of the entangled state given 
by a superposition of a spin-up electron with a spin-down muon, 
the spin state being given with respect to some choice of axis. 
What distinguishes the state space of a pair of spin-$\frac{1}{2}$ 
particles is the existence of a preferred singlet state 
$Z^{\alpha}$. This state is required to be self-conjugate polar 
with respect to the quadric in the sense that ${\bar Z}_{\alpha} 
= Q_{\alpha\beta}Z^{\beta}$. \par 

\section{Maximal entanglement} 

We are now in a position to present a more geometrical construction 
for the supremum of the cross-ratio $\kappa$ under the given 
constraint. Given the entangled state $\psi^{\alpha}$ we wish to 
find the state $X^{\alpha}\in Q$ that maximises the cross ratio 
\[ 
\kappa\ =\ \frac{(\psi^{\alpha}{\bar X}_{\alpha})
(X^{\beta}{\bar \psi}_{\beta})}{(\psi^{\alpha}
{\bar \psi}_{\alpha})(X^{\beta}{\bar X}_{\beta})} . 
\] 
In fact, suppose we define ${\bar\psi}^{\alpha}:=Q^{\alpha\beta}
{\bar\psi}_{\beta}$, the polar state of the complex conjugate 
hyperplane ${\bar\psi}_{\alpha}$. Then we can show that the 
states on $Q$ that are maximally and minimally distant to the 
given state $\psi^{\alpha}$ are collinear with $\psi^{\alpha}$, 
and are complex conjugate polar to one another in the sense 
that $\psi^{\alpha}$ has to be of the form 
\[ 
\psi^{\alpha}\ =\ pX^{\alpha}+qQ^{\alpha\beta}{\bar X}_{\beta} 
\] 
where $X^{\alpha}$ is the point on $Q$ closest to $\psi^{\alpha}$, 
so $|p|\geq|q|$. This can be verified, for example, by maximising 
$\kappa$ with 
respect to $X^{\alpha}$ subject to the constraint $Q_{\alpha\beta}
X^{\alpha}X^{\beta}=0$, using a Lagrange multiplier technique. 
Then if we define $\lambda=p/q$ it follows by a direct substitution 
that 
\[ 
\kappa\ =\ \frac{\lambda{\bar\lambda}}
{1+\lambda{\bar\lambda}}.
\] 
Since $\lambda{\bar\lambda}\geq1$, it follows, further, that 
$\frac{1}{2}\leq\kappa\leq1$. On the other hand, we can also 
verify by direct substitution that the invariant $\rho$ defined 
by 
\[ 
\rho\ =\ \frac{(Q_{\alpha\beta}\psi^{\alpha}\psi^{\beta})
({\bar Q}^{\gamma\delta}{\bar\psi}_{\gamma}{\bar\psi}_{\delta})}
{(\psi^{\gamma}{\bar\psi}_{\gamma})^{2}} , 
\] 
which is independent of the scale of $Q_{\alpha\beta}$, depends on 
$p$ and $q$ only through $\lambda$, and is given by the formula 
\[ 
\rho\ =\ \frac{4\lambda{\bar\lambda}}
{(1+\lambda{\bar\lambda})^{2}} . 
\] 
Then it is not difficult to see that $\kappa$ is indeed of the 
desired form $\kappa=\frac{1}{2}(1+\gamma)$ with 
\[ 
\gamma\ =\ \sqrt{1-\rho} . 
\] 
That establishes the the validity of the expression indicated 
earlier for the minimum distance 
\[ 
\delta\ =\ \cos^{-1}\sqrt{1-\rho} 
\] 
from the given state $\psi^{\alpha}$ to the quadric of 
disentanglement. \par 

The maximally entangled states are those for which $|\lambda|=1$, 
for which apart from an overall irrelevant scale factor 
$\psi^{\alpha}$ is thus necessarily of the form 
\[ 
\psi^{\alpha}\ =\ e^{{\rm i}\theta}X^{\alpha} + 
e^{-{\rm i}\theta}
Q^{\alpha\beta}{\bar X}_{\beta} . 
\] 
Such states are self-conjugate in the sense that 
${\bar\psi}_{\alpha}=Q_{\alpha\beta}\psi^{\beta}$. Conversely, given 
any disentangled state $X^{\alpha}$ we see that there exists a 
one-parameter family of maximally entangled states at a distance 
$\pi/2$ from it. This one-parameter family is given by the 
equatorial circle of the complex projective line obtained by joining 
$X^{\alpha}$ to the conjugate entangled state 
$Q^{\alpha\beta}{\bar X}_{\beta}$, to which $X^{\alpha}$ is 
orthogonal. \par 

Thus, for example, if $X^{AB}=\xi^{A}\eta^{B}$ is a disentangled 
state of two spin-$\frac{1}{2}$ particles, then we obtain the 
one-parameter family of maximally entangled states given by 
$\psi^{AB} = e^{{\rm i}\theta}\xi^{A}\eta^{B} + 
e^{-{\rm i}\theta}{\bar\xi}^{A}{\bar\eta}^{B}$ 
where ${\bar\xi}^{A}:=\epsilon^{AB}{\bar\xi}_{B}$ and 
${\bar\eta}^{B}:=\epsilon^{BA}{\bar\eta}_{A}$. For any value of 
$\theta$ these states are at a distance of $\pi/2$ from 
$X^{AB}$. \par

A special case of interest arises when $\eta^{B}={\bar\xi}^{B}$ 
and ${\bar\eta}^{A}=-\xi^{A}$. In that case, reverting to the 
notation of the previous section, we have $\psi^{AB} = 
e^{{\rm i}\theta} \psi^{A}{\bar\psi}^{B} - 
e^{-{\rm i}\theta}{\bar\psi}^{A}\psi^{B}$. Then for 
$\theta=0$ we obtain the spin 0 singlet state for which 
$\psi^{AB}\propto \epsilon^{AB}$; whereas for $\theta=\pi$ we 
get the $S_{z}=0$ spin 1 triplet state for which $\psi^{AB}
\propto\psi^{(A}{\bar\psi}^{B)}$. \par 

\begin{figure}[t] 
   \psfig{file=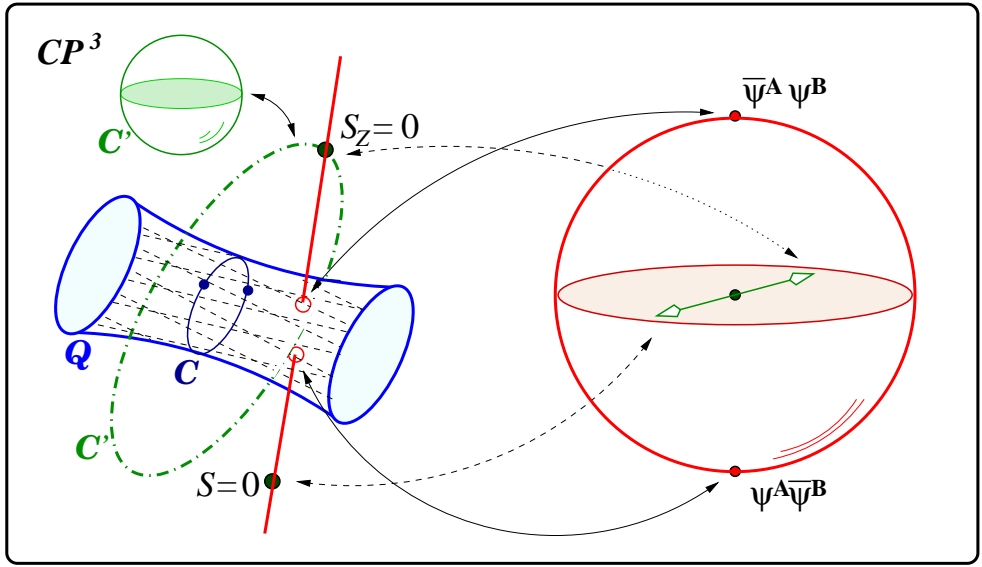,width=13cm,angle=0} 
\vskip .25cm 
 \caption{{\it Maximally entangled states}. Through any conjugate 
pair of disentangled states on $Q$ there exists a complex projective 
line containing the $S=0$ singlet and an $S_{z}=0$ triplet state 
for some choice of $z$-axis. The singlet and triplet states lie on 
an equatorial circle at a distance of $\pi/2$ from the 
disentangled states which are orthogonal to one another and thus 
lie on opposite poles. All the points on this equatorial circle are 
maximally entangled. The trajectory of the entangled triplet states 
corresponding to different spatial directions is a conic $C'$, 
which has a topology of a sphere. Hence the space of maximally 
entangled states has the structure of $S^{2}\times S^{1}$. 
} 
\end{figure} 

More generally, if $\psi^{\alpha}$ is an {\it arbitrary} maximally 
entangled state, then consider the conic ${\cal K}$ that arises 
when we intersect the plane ${\bar\psi}_{\alpha}$ with the 
quadric $Q$. This conic is conjugate self-polar in the sense that 
for any point $\pi^{\alpha}$ on ${\cal K}$ the complex conjugate 
plane ${\bar\pi}_{\alpha}$ is tangent to the quadric at a point 
${\bar\pi}^{\alpha}$ on ${\cal K}$. Now, suppose we consider the 
locus ${\cal L}$ of points generated by the intersection of the 
tangent lines to $\pi^{\alpha}$ and ${\bar\pi}^{\alpha}$ in the 
plane ${\bar\psi}_{\alpha}$ as we vary $\pi^{\alpha}$. For any 
point $P$ in ${\cal L}$ the join of that point with 
$\psi^{\alpha}$ intersects $Q$ in a pair of points $X^{\alpha}$ 
and ${\bar X}^{\alpha}$, both of which are at a distance 
$\delta=\pi/2$. By varying $P$ we obtain all points on $Q$ at 
a distance $\pi/2$ from $\psi^{\alpha}$. \par 

Finally, let us consider the case of sub-maximally entangled 
states. In this situation the relation between $\psi^{\alpha}$ 
and $X^{\alpha}$ is invertible, since providing $|\lambda|>1$ 
there exist complex numbers $r$ and $s$ such that 
\[ 
X^{\alpha}\ =\ r\psi^{\alpha}+sQ^{\alpha\beta}{\bar\psi}_{\beta}. 
\] 
We can solve this for the ratio $\mu=r/s$ by imposing the 
condition $Q_{\alpha\beta}X^{\alpha}X^{\beta}=0$, leading to 
the quadratic equation 
$\mu^{2}Q_{\alpha\beta}\psi^{\alpha}\psi^{\beta} + 2\mu 
\psi^{\alpha}{\bar\psi}_{\alpha} + {\bar Q}^{\alpha\beta} 
{\bar\psi}_{\alpha}{\bar\psi}_{\beta} = 0$, 
for which the roots are given by 
\[ 
\mu\ =\ \frac{-1\pm\sqrt{1-q{\bar q}}}{q} , 
\] 
where $q:=Q_{\alpha\beta}\psi^{\alpha}\psi^{\beta}/\psi^{\gamma}
{\bar\psi}_{\gamma}$. The positive root gives the point on $Q$ 
nearest to $\psi$, and the negative root gives the most distant 
disentangled state. We note that the terms here are so 
constructed that the solution for $X^{\alpha}$ is independent of 
the overall scale and phase of $\psi^{\alpha}$, as expected. \par 
  
\section{Schr\"odinger evolution} 

As the examples above indicate, the geometry of quantum mechanics 
is very rich, once specific physical systems are brought into play, 
even when there are only a few degrees of freedom. This picture can 
be further developed by consideration of the dynamics of a quantum 
system, which can be pictured as a vector field on the state 
manifold. Such a vector field generates a symmetry of the 
Fubini-Study geometry, i.e., an action of the projective 
unitary group. \par 

In the case of an $(n+1)$-dimensional Hilbert space, the state 
space is $CP^{n}$, which can be viewed as a real manifold 
${\sl\Gamma}$ of dimension $2n$, with a symmetry group of 
dimension $n(n+2)$, generated by a family of $n(n+2)$ Killing 
vector fields. A generic Killing field on ${\sl\Gamma}$ has $n+1$ 
fixed points, corresponding to eigenstates of the given 
Hamiltonian. \par 

In the case of a 2-dimensional 
Hilbert space, the state space is $CP^{1}$, and the 
specification of a Killing field selects out a pair of polar 
points on $S^{2}$, corresponding to energy eigenstates $E_{1}$ 
and $E_{2}$. The relevant symmetry is then given by a rigid 
rotational flow about this axis, the angular frequency being 
determined by Planck's formula $E_{2}-E_{1}=\hbar\omega$. 
In the case of the state space $CP^{n}$, the $n+1$ fixed points 
of a given Killing field are linked by a figure consisting of 
$\frac{1}{2}n(n+1)$ spheres, for which the fixed points act as 
polar points, in pairs. These spheres rotate 
respectively with angular frequencies 
\[ 
E_{i}-E_{j}\ =\ \hbar\omega_{ij}\ , 
\] 
where $E_{i}$ ($i=1,2,\cdots,n+1$) 
labels the energy of $i$th eigenstate. The dynamical 
trajectories in ${\sl\Gamma}$ are determined by the 
specification of the fixed points, along 
with the associated angular frequencies.
Even in the case of a simple spin 1 system, the 
geometry of the state space is intricate, given 
by a 4-dimensional manifold containing three 2-spheres 
touching one another at the poles, and spinning at 
three distinct frequencies. \par 

\begin{figure}[t] 
   \psfig{file=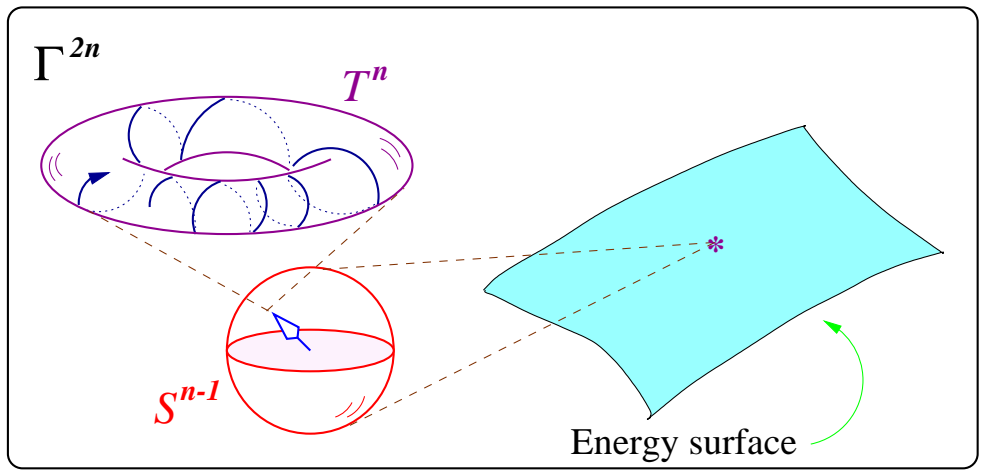,width=13cm,angle=0} 
\vskip .25cm 
 \caption{{\it Foliation of the state space and quantum dynamical 
trajectories}. The quantum state space is foliated by level 
surfaces of the expectation of the Hamiltonian operator. Each 
point on such a surface is 
parameterised by a family of $n$ phase variables, $n-1$ angular 
coordinates, and one energy variable. The Schr\"odinger evolution 
preserves the energy and the angular coordinates of a given initial 
state (point), and the resulting dynamical trajectory is thus 
confined to the $n$-torus $T^{n}$. If the ratios $E_{i}/E_{j}$ of 
the energy eigenvalues are irrational, then the trajectory on 
$T^{n}$ does not close.  
} 
\end{figure} 

If the frequencies are not commensurate in the sense of being 
rational multiples of each other, then the Killing orbits do 
not close except on the three special spheres, and the generic 
dynamical trajectory, starting from some initial point in the 
state space, is doomed to evolve to eternity without ever 
returning to its origin. More specifically, we can view the 
quantum state space $CP^{n}$ as a real manifold ${\sl\Gamma}^{2n}$ 
of dimension $2n$. Then, we can consider a foliation of 
${\sl\Gamma}^{2n}$ by level surfaces of constant energies. The 
specification of the energy thus determines one of the $2n$ real 
coordinates on the state space. The remaining $2n-1$ coordinates 
can be identified (Brody $\&$ Hughston 1999b) with $n-1$ angular 
coordinates and $n$ phase variables. In other words, each energy 
surface has the structure of a product space of an $n$-torus and 
an ($n-1$)-sphere, i.e., $T^{n}\times S^{n-1}$. Given an initial 
point on one of the energy surfaces, the dynamics induced by the 
Schr\"odinger evolution will confine that point to the torus that 
contains the initial state. In other words, the angular variables 
of the initial state remain unchanged under the action of the 
projective unitary group induced by the given quantum Hamiltonian. 
It follows, therefore, that the Schr\"odinger evolution is at best 
merely quasiergodic on the energy surface. \par 

\section{Quantum Hamiltonian dynamics} 

This line of argument can be taken further by studying the quantum 
trajectories on ${\sl\Gamma}$ by use of differential geometry. When 
viewed as a real manifold, the state space is endowed with a 
Riemannian structure, given by a positive definite symmetric metric 
$g_{ab}$, a symplectic structure, given by an antisymmetric tensor 
$\Omega_{ab}$, and a complex structure, given by a tensor 
$J^{a}_{\ b}$, satisfying 
\[ 
J^{a}_{\ c}J^{c}_{\ b}\ =\ -\delta^{a}_{\ b}\ . 
\] 
These structures are required to be {\it compatible} in the 
sense that $\Omega_{ab}=g_{ac}J^{c}_{\ b}$ and $\nabla_{a} 
J^{b}_{\ c}=0$, where $\nabla_{a}$ is the covariant derivative 
associated with $g_{ab}$. Here we use Roman indices ($a,b,\cdots$) 
for tensorial operations in the tangent space of ${\sl\Gamma}$. 
The compatibility of $g_{ab}$, 
$\Omega_{ab}$, and $J^{a}_{\ b}$ makes ${\sl\Gamma}$ a K\"ahler 
manifold. For some purposes it suffices to assume that the metric 
$g_{ab}$ and the symplectic structure $\Omega_{ab}$ are only 
{\it weakly} nondegenerate in the sense that for any vector fields 
$\xi^{a},\eta^{a}$ on ${\sl\Gamma}$, $\xi^{a}g_{ab}=0$ implies 
$\xi^{a}=0$ and $\eta^{a}\Omega_{ab}=0$ implies $\eta^{a}=0$. 
The fact that in infinite dimensions the Fubini-Study metric and 
symplectic structure are {\it strongly} nondegenerate (Marsden 
$\&$ Ratiu 1999) means that many of the geometrical constructions 
carried out in finite dimensions carry through to the general 
quantum phase space.  \par 

The additional ingredient required for the specification of the 
dynamics is a Hamiltonian function $H(x)$ on ${\sl\Gamma}$. Then 
the general dynamical trajectories on ${\sl\Gamma}$ 
are given by 
\[ 
\frac{1}{2}\hbar \Omega_{ab}dx^{b}\ =\ \nabla_{a}H dt\ . 
\] 
The Schr\"odinger trajectories on ${\sl\Gamma}$ are given by a 
subclass of the general Hamiltonian trajectories, namely, those 
for which the Hamiltonian function $H(x)$ is of the special form 
\[ 
H(x)\ =\ 
\frac{{\bar\psi}_{\alpha}(x)H^{\alpha}_{\beta}\psi^{\beta}(x)}
{{\bar\psi}_{\gamma}(x)\psi^{\gamma}(x)}\ . 
\] 
Here, $\psi^{\alpha}(x)$ denote homogeneous coordinates for the 
corresponding point $x$ in the projective Hilbert space. Thus for 
a Schr\"odinger trajectory, $H(x)$ is 
the expectation of the Hamiltonian operator in the pure state to 
which the point $x$ corresponds. In contrast with classical 
mechanics, where the phase space often has an 
interpretation in terms of position and momentum variables, in 
quantum mechanics the points in phase space correspond to pure 
quantum states. \par 

Quantum observables are intimately related to the metrical geometry 
of ${\sl\Gamma}$. The distinguishing feature of a quantum 
Hamiltonian function $H(x)$ is that the associated symplectic 
gradient flow $\xi^{a}=dx^{a}/dt$ is a 
Killing field, i.e., $\nabla_{(a}\xi_{b)}=0$. Indeed all Killing 
fields on ${\sl\Gamma}$ arise in this way through quantum 
observables. The Killing fields generate the symmetries 
of the Fubini-Study metric $g_{ab}$. \par 

In the case of finite dimensions, we can say more about the 
quantum observables that generate isometries on the Fubini-Study 
manifold. If $H(x)$ is a linear observable function, then in finite 
dimensions it is necessarily defined globally on ${\sl\Gamma}$. 
In fact, one can show that such functions correspond to global 
solutions of the characteristic equation 
\[ 
\nabla^{2}H\ =\ (n+1)({\bar H}-H)\ , 
\] 
where $\nabla^{2}$ is the Laplace-Beltrami operator on 
${\sl\Gamma}$, ${\bar H}=H^{\alpha}_{\alpha}/(n+1)$ is the uniform 
average of the eigenvalues of $H^{\alpha}_{\beta}$, and $2n$ is the 
real dimension of ${\sl\Gamma}$. 
Conversely, if we are given a Killing field $\xi^{a}$, the 
corresponding observable function $H(x)$ can be recovered, up to an 
additive constant, via the relation 
\[ 
\hbar\Omega^{ab}\nabla_{a}\xi_{b}\ =\ 2(n+1)(H-{\bar H})\ , 
\] 
which follows directly from the characteristic equation if we make 
use of the fact that $\hbar\xi_{a}=2J^{b}_{\ a}\nabla_{b}H$. We 
note, incidentally, that in the Kibble-Weinberg theory, a general 
nonlinear quantum observable is a function on ${\sl\Gamma}$ such 
that the characteristic equation is {\it not} satisfied. As a 
consequence, the corresponding symplectic gradient flow is no 
longer a Killing field. \par 

\section{Uncertainty relations} 

The metrical geometry of ${\sl\Gamma}$ also determines the 
statistical properties of observables. For example, in the pure 
state $x$ the squared uncertainty (variance) of an observable 
represented by the function $F(x)$ is $(\Delta F)^{2} 
= g_{ab}F^{a}F^{b}$ where $F^{a}$ is the unique gradient vector 
field satisfying $g_{ab}F^{b}=\nabla_{a}F$. This leads to the 
following interpretation of quantum 
mechanical uncertainty. We foliate ${\sl\Gamma}$ with 
surfaces given by level values of $F(x)$. Through a given 
pure state $x$ there is a unique such surface, and the 
uncertainty $\Delta F$ is the length of the gradient vector 
to that surface at $x$. 
The observables $F(x)$ and $G(x)$ are incompatible if their Poisson 
bracket $[F,G]=\Omega_{ab}F^{a}G^{b}$ is nonvanishing. 
In that case the Heisenberg uncertainty relation 
\[ 
(\Delta F)^{2}(\Delta G)^{2}\ \geq\ \frac{1}{4} 
\left| [F,G]\right| ^{2} 
\] 
follows directly as a consequence of the geometric inequality 
\[ 
(g_{ab}F^{a}F^{b})(g_{ab}G^{a}G^{b}) \ \geq\ 
(g_{ab}F^{a}G^{b})^{2} + \frac{1}{4}
(\Omega_{ab}F^{a}G^{b})^{2}\ , 
\] 
if we omit the first term in the right hand side. This inequality 
holds for any vector fields $F^{a}$ and $G^{a}$ on a K\"ahler 
manifold. Note that the omitted term $g_{ab}F^{a}G^{b}$ gives 
rise to the anticommutator of the observables $F$ and $G$. \par 

\begin{figure}[b] 
   \psfig{file=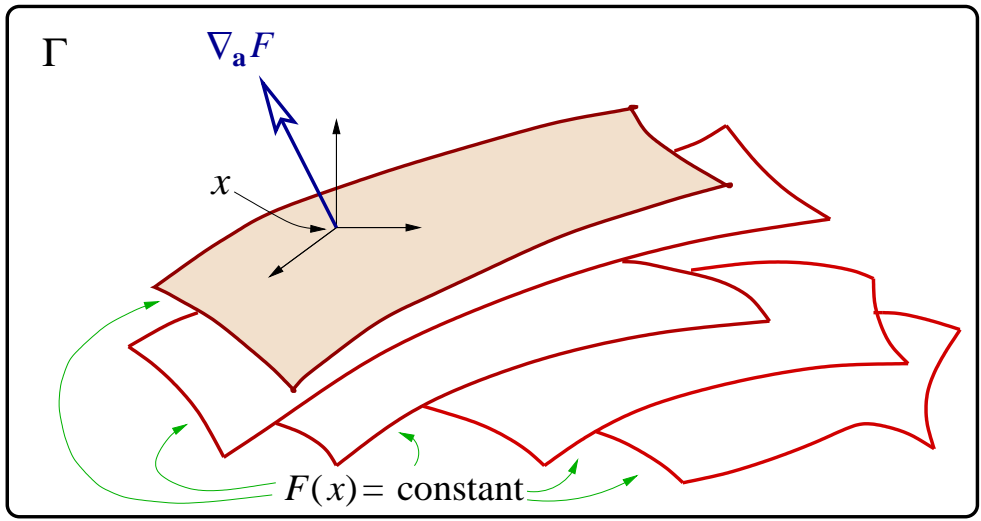,width=13cm,angle=0} 
\vskip .25cm 
 \caption{{\it The observable uncertainty}. The quantum phase 
space ${\sl\Gamma}$ is foliated by level surface of the 
function $F(x)$. The quantum uncertainty in the corresponding 
observable, in the pure state $x$, is given by the magnitude of 
the gradient of $F(x)$ at that point.  
} 
\end{figure} 

In the case of a pair of 
canonically conjugate observables $P(x)$ and $Q(x)$ satisfying 
$[P,Q]=\hbar$, we can expand the gradient to the surfaces of constant 
$Q(x)$ in a suitable basis to obtain a series of {\it generalised 
Heisenberg relations} (Brody $\&$ Hughston 1996, 1997, 1998a,b), an 
example of which is 
\[ 
(\Delta P)^{2}(\Delta Q)^{2}\ \geq\ \frac{1}{4}\hbar^{2} \left( 
1+\frac{(\mu_{4}(P)-3\mu_{2}(P)^{2})^{2}}
{\mu_{6}(P)\mu_{2}(P)-\mu_{4}(P)^{2}}\right) \ , 
\] 
where $\mu_{k}(P)=\langle(P-\langle P\rangle)^{k}\rangle$ is the 
$k$th central moment of the observable $P$ in the state $x$. This 
inequality has the following statistical interpretation. Suppose 
that we are given an unknown quantum state of a particle, 
parameterised by its position $q$, and that we wish to estimate 
the position of the particle by a suitable measurement. The 
observable function corresponding to the parameter $q$ is then 
given by $Q$, and the statistical estimation of $q$ via measurement 
on $Q$ gives rise to an inevitable variance lower bound, expressed 
in terms of a certain combination of the moments $\mu_{k}(P)$ of 
the momentum distribution associated with the given state. 
Likewise, if we consider momentum estimation, then the 
corresponding variance lower bound is given by the moments of the 
position $Q$. \par 

\section{Geometric phases} 

\begin{figure}[b] 
   \psfig{file=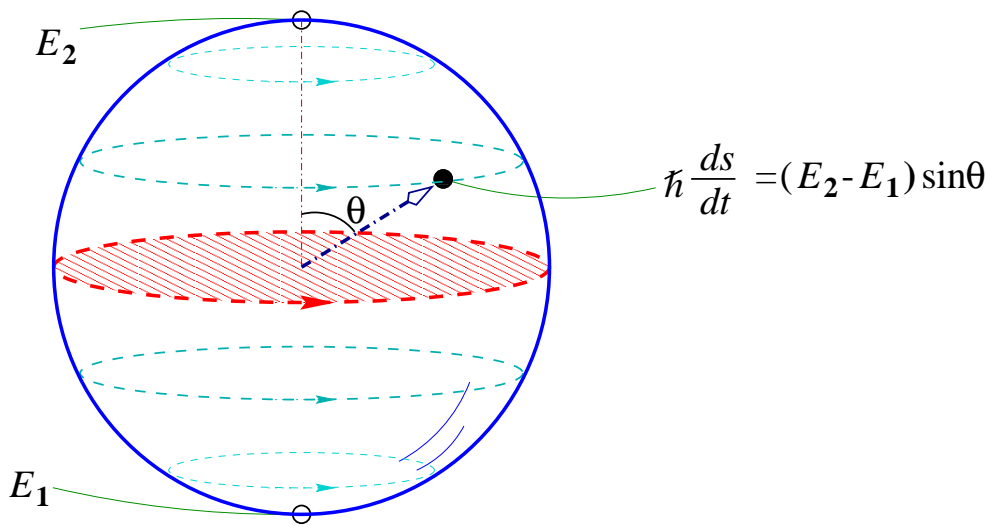,width=13cm,angle=0} 
\vskip .25cm 
 \caption{{\it The Anandan-Aharonov relation}. The quantum 
evolution of a 2-state system corresponds to the rigid rotation 
of a 2-sphere with angular frequency $\hbar\omega=E_{2}-E_{1}$. 
The speed of the trajectory is greatest at the equator, which 
consists of states of maximal energy uncertainty. 
} 
\end{figure} 
 
An interesting interplay between the quantum dynamical trajectories 
and the uncertainty relations was pointed out by Aharonov $\&$ 
Anandan (1990). In particular, it follows from the 
projective Schr\"odinger equation $\hbar\Omega_{ab}dx^{b} = 
2\nabla_{a}H dt$ and the expression for the line element $ds^{2}= 
g_{ab}dx^{a}dx^{b}$ that the `speed' in the state space 
${\sl\Gamma}$ along the dynamical trajectory at the point $x$ is 
given by 
\[ 
\hbar \frac{ds}{dt}\ =\ 2\Delta H\ , 
\] 
where $\Delta H$ is the energy uncertainty in the given state. 
For example, in the case of a 2-state system with eigenstates 
at the poles of a 2-sphere, the quantum evolution corresponds to 
a rigid rotation of the sphere, with constant angular frequency, 
for which the speed is greatest at the equator, 
corresponding to states of maximum uncertainty. 

\begin{figure}[b] 
  \psfig{file=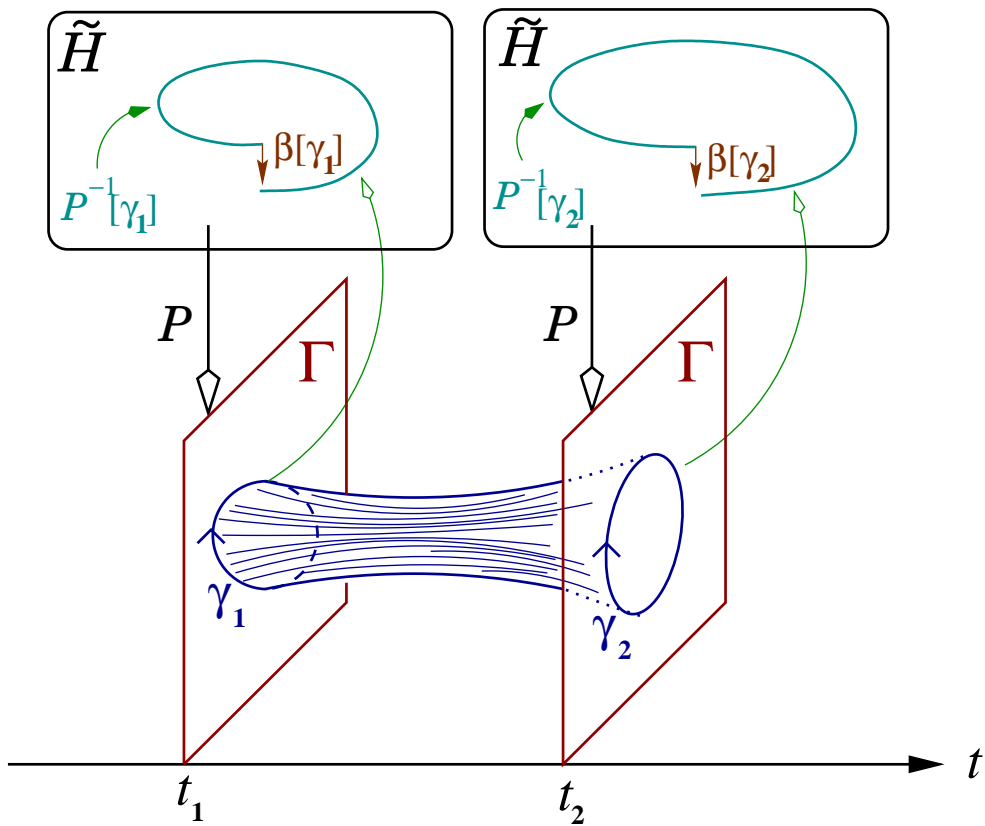,width=13cm,angle=0} 
\vskip .25cm 
 \caption{{\it The horizontal lift of a quantum trajectory and 
Poincare's invariant integral}. The Berry phase $\beta[\gamma]$ 
associated with a general cyclic trajectory $\gamma$ in the 
quantum phase space ${\sl\Gamma}$ is given by the integral of the 
symplectic form $\Omega_{ab}$ over a 2-surface $\Sigma$ spanning 
$\gamma$. This integral measures the phase change that develops in 
the horizontal lift of $\gamma$ to the corresponding path 
${\cal P}^{-1}[\gamma]$ in the Hilbert bundle ${\tilde{\cal H}}$ 
over ${\sl\Gamma}$. If the cyclic trajectory subsequently evolves 
unitarily in time, then $\beta[\gamma]$ is the quantum mechanical 
integral invariant of Poincare. As a consequence, we have 
$\beta[\gamma_{1}] = \beta[\gamma_{2}]$. This result is valid 
even if we relax the unitarity condition and consider nonlinear 
dynamics of the Kibble-Weinberg type. 
} 
\end{figure} 

This result is related to properties of the {\it geometric phase} 
introduced by Berry and subsequently applied in many situations 
(Simon 1983; Berry 1984; Uhlmann 1986, see also Shapere $\&$ 
Wilczek 1989). Consider a closed path $\gamma$ in 
the quantum phase space. If $\gamma$ is a standard dynamical 
trajectory, then it corresponds to a closed Killing orbit, but 
we shall allow for the possibility of more general paths, e.g., 
as might be generated by a time-dependent Hamiltonian operator. 
The geometric phase associated with such a cyclic evolution is 
given by 
\[ 
\beta[\gamma]\ =\ \int_{\Sigma}\Omega_{ab} 
dx^{a}\wedge dx^{b}\ , 
\] 
where $\Sigma$ is any real 2-surface in ${\sl\Gamma}$ such that 
$\gamma=\partial\Sigma$. Owing to the relation 
$\nabla_{a}\Omega_{bc}=0$, it follows from Stokes' theorem that 
the value of $\beta[\gamma]$ is independent of the choice of 
surface $\Sigma$ spanning the loop $\gamma$, and can be given the 
following interpretation. \par  

The punctured Hilbert space ${\tilde{\cal H}}={\cal H}-\{0\}$, 
obtained by deleting the origin, is a fibre bundle over 
${\sl\Gamma}$. Therefore, given a 
trajectory $\gamma$ in ${\sl\Gamma}$, we can form a corresponding 
trajectory ${\cal P}^{-1}[\gamma]$ in ${\tilde{\cal H}}$, called 
horizontal lift of $\gamma$. This is obtained by solving the 
{\it modified} Schr\"odinger equation 
\[ 
{\rm i}\hbar\frac{\partial\psi^{\alpha}}{\partial t}\ =\ 
(H^{\alpha}_{\beta}-{\rm E}[H]\delta^{\alpha}_{\beta})
\psi^{\beta}, 
\] 
where ${\rm E}[H]$ is the expectation of the Hamiltonian in the 
state $\psi^{\alpha}$. Despite its nonlinearity, the modified 
Schr\"odinger equation is physically natural inasmuch as its 
stationary states are energy eigenstates. In this connection, it 
is worth drawing attention to the fact that in the case of the 
modified Schr\"odinger dynamics, the time independent 
Schr\"odinger equation 
\[ 
H^{\alpha}_{\beta}\psi^{\beta} \ =\ {\rm E}[H]\psi^{\alpha} 
\] 
follows directly from the stationary state requirement, without 
the introduction of the so-called correspondence principle 
${\rm E}[H]\leftrightarrow{\rm i}\hbar\partial_{t}$. \par 

The horizontal lift is characterised by the condition that the 
tangent to the curve ${\cal P}^{-1}[\gamma]$ in ${\tilde{\cal H}}$, 
given by $\partial\psi^{\alpha}/\partial t$, is orthogonal to the 
fibre direction $\psi^{\alpha}$, so we have ${\bar\psi}_{\alpha}
\partial\psi^{\alpha}/\partial t=0$. \par 

In the case of a closed loop $\gamma$, $\beta[\gamma]$ measures 
the phase change in $\psi^{\alpha}$ over the corresponding 
circuit in ${\cal P}^{-1}[\gamma]$. If the given loop 
$\gamma$ in ${\sl\Gamma}$ subsequently evolves in time, then 
$\beta[\gamma]$ is a quantum mechanical analogue of the Poincare 
integral invariant (cf. Arnold 1989). We note, incidentally, that 
the notion of geometric phase discussed here also applies to 
nonlinear quantum mechanics, for which the Hamiltonian 
$H(x)$ does not satisfy the characteristic equation for linear 
observables. \par 
 
\section{Mixed states} 

Phase space geometry sheds some interesting 
light on the role of probability in quantum mechanics. There are 
at least two situations where probability distributions on the 
state manifold ${\sl\Gamma}$ have to be considered. One is 
in the description of the statistical properties of a measurement 
outcome; the other is in quantum statistical mechanics. \par 

In both cases, the state of the system can be characterised by a 
probability density function $\rho(x)$ on ${\sl\Gamma}$, in terms 
of which the expectation of any function $F(x)$ on ${\sl\Gamma}$ 
can be written 
\[ 
{\rm E}[F]\ =\ \int_{\sl\Gamma}\rho(x)F(x)dx . 
\] 
We think of $F(x)$ as representing the expectation of the 
corresponding observable, {\it conditional} on the pure state $x$. 
Then ${\rm E}[F]$ is the {\it unconditional} expectation, where 
we average $F(x)$ over the pure states, weighting with the density 
$\rho(x)$. A pure state arises if $\rho(x)$ is a $\delta$-function 
concentrated on a point in ${\sl\Gamma}$. 
Consider the example of a measurement where initially 
the system is in a pure state $X$, and the observable has a 
finite number of eigenstates, as in the case of a spin-1 
system when we measure the spin along an axis. The result 
of this measurement is one of the three spin eigenstates, 
and these arise with probabilities determined by the Fubini-Study 
distance. Thus the density function $\rho(x)$ for the state 
of the system after a measurement is given by a sum of three 
$\delta$-functions, concentrated at the eigenstates, 
weighted by these probabilities. \par 

In the case of a quantum observable, the unconditional variance 
of $F(x)$ in a general mixed state $\rho(x)$ is given by 
\[ 
{\rm V}[F]\ =\ \int_{\sl\Gamma} \rho(x)(F(x)-{\rm E}[F])^{2}dx 
+ \int_{\sl\Gamma}\rho(x)(g_{ab}F^{a}F^{b})^{2}dx\ . 
\] 
A further simplification emerges by virtue of the special form 
of a linear observable, for which we have ${\rm E}[F] = 
\rho^{\alpha}_{\beta}F^{\beta}_{\alpha}$, where 
\[ 
\rho^{\alpha}_{\beta}\ =\ \int_{\sl\Gamma}\rho(x) 
\frac{{\bar\psi}_{\beta}(x)\psi^{\alpha}(x)}
{{\bar\psi}_{\gamma}(x)\psi^{\gamma}(x)}dx 
\] 
is the {\it density matrix} associated with $\rho(x)$. For ordinary 
linear quantum mechanics it suffices to consider the density matrix 
alone, since all statistical quantities calculated with $\rho(x)$ 
reduce to expressions involving $\rho^{\alpha}_{\beta}$. 
Therefore, for certain purposes we can regard 
$\rho^{\alpha}_{\beta}$ itself as representing the state of the 
system. \par 

One should bear in mind, however, that the density matrix 
$\rho^{\alpha}_{\beta}$, which is the lowest moment of the 
projection operator in the state $\rho(x)$, does not in general 
contain all the information of the system when we are dealing 
with nonlinear observables. This follows from the fact that the 
information of a generic state $\rho(x)$ is contained in the set 
of {\it all} the moments (cf. Brody $\&$ Hughston 1999c). More 
specifically, in the case of a nonlinear observable, we must 
consider a general state $\rho(x)$, pure or mixed, because the 
density matrix is not sufficient to take the expectation of such 
an observable. Some specific examples of nonlinear observables 
have been studied by Weinberg (1989a,b). The entanglement measure 
$\delta$ introduced in \S 8 provides another explicit example of 
a nonlinear observable arising in a natural context. Exclusive 
consideration of the density matrix in a nonlinear setting can 
lead to paradoxical and apparently nonphysical conclusions, such 
as the possibility of superluminal EPR communication (cf. Gisin 
1989; Polchinski 1991). \par 

Given a general state $\rho(x)$ and a Hamiltonian $H(x)$, the 
evolution of $\rho(x)$ is governed by the Liouville equation, 
\[ 
\frac{1}{2}\hbar{\dot\rho}(x)\ =\ 
\Omega^{ab}\nabla_{a}\rho\nabla_{b}H \ , 
\] 
where the Poisson bracket between $\rho(x)$ and $H(x)$ 
is determined by the symplectic structure $\Omega_{ab}$ on 
${\sl\Gamma}$. In the case where the Hamiltonian is a linear 
quantum observable, the Liouville 
equation is equivalent to the standard Schr\"odinger dynamics 
associated with a mixed state $\rho(x)$. On the other hand, if 
the Hamiltonian is a nonlinear observable, then the Liouville 
equation no longer corresponds to a linear Schr\"odinger 
evolution. \par 

It is interesting to note, nevertheless, that, contrary to what 
has been argued in literature (cf. Peres 1989), in the case of 
nonlinear quantum mechanics of the Kibble-Weinberg type, the 
quantum entropy 
\[ 
S(\rho)\ =\ -\int_{\sl\Gamma}\rho(x)\ln\rho(x) dx 
\] 
associated with a general mixed state $\rho(x)$ remains constant 
in time. This follows as a direct consequence of the Liouville 
equation for $\rho(x)$. \par 

More generally, the definition of entropy and equilibrium in 
quantum statistical mechanics brings up important conceptual 
issues, since, like the quantum measurement problem, it involves 
the interface of microscopic and macroscopic physics. There is 
also a profound relationship to fundamental issues in probability 
theory. Suppose we consider a quantum system 
characterised by a state space ${\sl\Gamma}$ and a Hamiltonian 
function $H(x)$ with discrete, possibly degenerate energy levels 
$E_{j}$ $(j=1,2,\cdots,N)$. Let us write $\delta_{j}(x)$ for a 
normalised $\delta$-function on ${\sl\Gamma}$ concentrated on 
the pure state $x_{j}$ with energy $E_{j}$. Thus, $x_{j}$ is the 
$j$th energy eigenstate. Then if the quantum system is in 
equilibrium with a heat-bath at inverse temperature $\beta=1/kT$, 
the state of the system is 
\[ 
\rho(x)\ =\ \frac{\sum_{j}\exp(-\beta E_{j})\delta_{j}(x)}
{Z(\beta)}\ , 
\] 
where $Z(\beta)=\sum_{j}\exp(-\beta E_{j})$ is the partition 
function. This is the canonical distribution of quantum 
statistical mechanics, characterised by a Gibbs distribution 
concentrated on the energy eigenstates with Boltzmann weights 
$\exp(-\beta E_{j})/Z(\beta)$. The standard canonical 
density matrix associated with this distribution is 
$\rho^{\alpha}_{\beta}=\exp(-\beta H^{\alpha}_{\beta})
/Z(\beta)$, which is clearly independent of the phase and scale 
of the underlying energy eigenvectors, and thus can be regarded 
as belonging to the geometry of ${\sl\Gamma}$. \par 

\section{Quantum theory and beyond} 

There is a paradox at the heart of statistical mechanics, 
related to the fact that there are many distinct probability 
distributions on ${\sl\Gamma}$ that give rise to the canonical 
density matrix. A natural question to ask, therefore, is whether 
there exists a `preferred' density function on ${\sl\Gamma}$ for 
the canonical ensemble. In the case of classical mechanics, the 
maximum entropy argument `selects' a preferred distribution. The 
problem here is that when applied to quantum mechanics, this 
argument leads to a quantum canonical ensemble characterised by 
the distribution 
\[ 
\rho(x)\ =\ \frac{\exp(-\beta H(x))}
{\int_{\sl\Gamma}\exp(-\beta H(x))dx}\ , 
\] 
rather than the system of weighted 
$\delta$-functions concentrated on energy eigenstates indicated 
earlier (Brody $\&$ Hughston 1998c, 1999a). However, the maximum 
entropy ensemble on 
${\sl\Gamma}$ leads to a density matrix different from the 
canonical density matrix. This apparent contradiction may 
ultimately be resolved by a more refined consideration of the 
available empirical evidence. The key point here is that even if 
the macroscopic energy of a substance in thermal equilibrium 
with a fixed heat bath is specified, there is no known principle 
that requires the individual subconstituents of that substance to 
be in energy eigenstates.  \par 

One further reason for the consideration of general probability 
distributions on ${\sl\Gamma}$ is that such states are necessary 
for an account of the statistical properties of observables in 
nonlinear quantum mechanical systems. These systems were given a 
general characterisation in terms of quantum phase space geometry 
by Kibble (1978, 1979), who observed that if we 
keep the phase space ${\sl\Gamma}$ of quantum mechanics, along 
with the Fubini-Study metric and the associated symplectic 
structure, but extend the category of observables to include 
general functions on ${\sl\Gamma}$, then the corresponding 
nonlinear Schr\"odinger dynamics can still be expressed in 
Hamiltonian form, i.e., $\hbar\Omega_{ab}dx^{b} = 2\nabla_{a}H 
dt$. Here $H(x)$ represents a general nonlinear functional of the 
wave function, not necessarily given by the expectation of a 
linear operator. \par 

An example of such a nonlinear evolution is given by the 
Newton-Schr\"odinger equation. Consider a quantum system of 
self-gravitating particles, described by the Schr\"odinger 
equation in ${\bf R}^{3}$ with a potential $\phi({\bf x})$, 
as described earlier, where $\phi({\bf x})$ is the gravitational 
potential due to the probable mass distribution of the quantum 
system, given by the Poisson equation 
\[ 
\nabla^{2}\phi({\bf x})\ =\ 4\pi m p({\bf x}) , 
\] 
where $p({\bf x})={\bar\psi}({\bf x})\psi({\bf x})/\int
{\bar\psi}({\bf x})\psi({\bf x})d^{3}x$. Because the 
potential depends on $\psi({\bf x})$, the resulting Schr\"odinger 
equation is nonlinear. As another example of nonlinear dynamics we 
might envisage a modification of the Schr\"odinger equation that 
would tend to drive an initially entangled system towards a 
state of disentanglement. \par 

The general features of nonlinear quantum dynamics have been 
studied by a number of authors (Kibble $\&$ Randjbra-Daemi 1980; 
Weinberg 1989a,b,c; Peres 1989; Gisin 1989; Polchinski 1991; 
Gibbons 1992; Percival 1994), and it is both surprising and 
gratifying in this context how naturally geometric quantum 
mechanics can be adapted to so many aspects of the nonlinear 
regime. This suggests that the geometric approach may eventually 
be useful in solving some of the key open problems in quantum 
theory, e.g., a clear understanding of the process of state 
reduction and a proper integration of the theory with gravitation 
(Einstein et al. 1935; Wheeler $\&$ Zurek 1983; Bell 1987). \par

\begin{acknowledgments} 
The authors wish to express their gratitude to E.J. Brody, 
T.R. Field, G.W. Gibbons, L.P. Horwitz, T.W.B. Kibble, B.K. 
Meister, R. Penrose, S. Popescu, and R.F. Streater for 
stimulating discussions. DCB acknowledges PPARC and The Royal 
Society for financial support. \par 
\end{acknowledgments} 


\end{document}